\documentclass[aps, prx, reprint, superscriptaddress]{revtex4-2}
\renewcommand{\figurename}{Figure}

\usepackage{array}
\usepackage[dvips]{graphicx}
\usepackage{amsmath,amssymb,amsthm,mathrsfs,amsfonts,dsfont,amscd,keyval}
\usepackage{mathtools,mathrsfs}
\usepackage{yfonts} 
\usepackage{ytableau}
\usepackage{subfigure, epsfig}
\usepackage{braket}
\usepackage{bm}
\usepackage{fancyhdr}
\usepackage{enumerate}

\usepackage{hyperref}
\usepackage{tabularx}
\usepackage{times}
\hypersetup{colorlinks=true, linkcolor=purple, citecolor=purple, urlcolor=black}
\usepackage{float}
\usepackage{tensor}
\usepackage{multirow}
\usepackage[normalem]{ulem}  
\usepackage{physics}

\usepackage{xcolor}
\usepackage{listings}
\usepackage{tcolorbox}
\usepackage{fancyvrb}
\usepackage{booktabs}
\usepackage{enumitem} 

\lstdefinestyle{quantumcode}{
    basicstyle=\ttfamily\small\color{white},
    backgroundcolor=\color{black!90},
    numbers=left,
    numberstyle=\tiny\color{gray},
    numbersep=8pt,
    frame=tb,
    framerule=0.5pt,
    rulecolor=\color{gray},
    tabsize=4,
    breaklines=true,
    postbreak=\mbox{\textcolor{green}{$\hookrightarrow$}\space},
    keywordstyle=\color{green!70!black},
    commentstyle=\color{cyan},
    stringstyle=\color{white},
    morekeywords={QuantumCircuit, append, h, measure, range}, 
    literate=* 
        {:}{{\textcolor{white}{:}}}{1} 
        {=}{{\textcolor{white}{=}}}{1}
        {(}{{\textcolor{white}{(}}}{1}
        {)}{{\textcolor{white}{)}}}{1}
        {*}{{\textcolor{white}{*}}}{1}
}

\newtheorem{conclusion}{Conclusion}[section]

\newcommand{\ie}{\textit{i.e.}}
\newcommand{\eg}{\textit{e.g.}}

\theoremstyle{definition}

\theoremstyle{plain}

\theoremstyle{remark}
\newtheorem*{remark}{Remark}
\newcommand{\comments}[1]{}

\makeatletter
\def\l@subsubsection#1#2{} 
\makeatother

\usepackage[T1]{fontenc}
\usepackage[utf8]{inputenc}
\usepackage{lmodern}
\usepackage{microtype}
\usepackage{bookmark}

\begin{document}

\let\oldaddcontentsline\addcontentsline

\renewcommand{\addcontentsline}[3]{}

\title{Symbolic Analysis of Grover Search Algorithm via Chain-of-Thought Reasoning and Quantum-Native Tokenization}

\author{Min Chen}
\affiliation{Department of Computer Science, The University of Pittsburgh, Pittsburgh, PA 15260, USA}

\author{Jinglei Cheng}
\affiliation{Department of Computer Science, The University of Pittsburgh, Pittsburgh, PA 15260, USA}

\author{Pingzhi Li}
\affiliation{Department of Computer Science, The University of North Carolina at Chapel Hill, Chapel Hill, NC 27599, USA}

\author{Haoran Wang}
\affiliation{Department of Computer Science, The University of North Carolina at Chapel Hill, Chapel Hill, NC 27599, USA}

\author{Tianlong Chen{$^\ddagger$}}
\affiliation{Department of Computer Science, The University of North Carolina at Chapel Hill, Chapel Hill, NC 27599, USA}

\author{Junyu Liu{$^\ddagger$}}
\affiliation{Department of Computer Science, The University of Pittsburgh, Pittsburgh, PA 15260, USA}

\date{Dated: November 11, 2024}

\date{\today}

\maketitle

\vfill
\begin{flushleft}
{$^{\ddagger}$Co-corresponding authors.}\\
\href{mailto:junyuliu@pitt.edu}{junyuliu@pitt.edu},
\href{mailto:tianlong@cs.unc.edu}{tianlong@cs.unc.edu}
\end{flushleft}

\section*{Abstract}

\noindent{\bf
Understanding the high-level conceptual structure of quantum algorithms from their low-level circuit representations is a critical task for verification, debugging, and education. While traditional numerical simulators can calculate output probabilities, they do not explicitly surface the underlying algorithmic logic, such as the function of an oracle or embedded symmetries. In this work, we shift the focus from numerical simulation to \textbf{symbolic analysis}, investigating whether Large Language Models (LLMs) can automatically \textbf{interpret} quantum circuits and articulate their logic in a human-readable format. We introduce GroverGPT+, a model that leverages Chain-of-Thought reasoning and quantum-native tokenization to analyze Grover's search algorithm. We use Grover's algorithm as a controlled testbed, as its well-defined analytical properties allow for rigorous verification of the model's reasoning process. Our primary finding is that GroverGPT+ successfully identifies the oracle and its marked states directly from circuit representations. The model's key output is not a final probability, but a structured, \textbf{interpretable reasoning trace} that mirrors human expert analysis, effectively translating procedural circuit steps into conceptual insights. Furthermore, we establish a structured benchmark for this symbolic analysis task and explore its empirical extrapolation describing the model's performance as the number of qubits increases. These findings position LLMs as powerful tools for automated quantum algorithm analysis and verification. More fundamentally, this work offers a first step towards using such models as scientific probes, suggesting that an algorithm's ``learnability" by a classical model can provide a new, complementary perspective on its conceptual complexity, a topic of core interest to quantum information science.}

\section{Introduction}

Quantum computing has emerged as a transformative paradigm, with algorithms like Shor's \citep{shor1999polynomial} and Grover's \citep{10.1145/237814.237866} demonstrating profound theoretical advantages over classical counterparts \citep{nielsen2010quantum}. However, a significant gap exists between the low-level procedural descriptions of quantum algorithms, such as Quantum Assembly Language (QASM) code \citep{cross2022openqasm}, and the high-level conceptual understanding required for their design, verification, and debugging. While traditional numerical simulators are essential tools for calculating the evolution of state vectors \citep{pednault2019leveraging,pan2021simulating,chundury2024diaq}, they are ``semantically blind": they output final probabilities but do not explicitly surface the underlying algorithmic logic, such as the function of an oracle or embedded symmetries within the circuit. This semantic gap presents a bottleneck, motivating the development of new tools that can automatically interpret and reason about the structure of quantum algorithms.

Recently, Large Language Models (LLMs) have demonstrated remarkable capabilities in bridging such semantic gaps in various domains, from code generation to complex scientific reasoning \citep{liu2024deepseek,wei2023chainofthoughtpromptingelicitsreasoning}. Their ability to process and generate structured, human-readable text makes them prime candidates for a task that moves beyond numerical calculation towards conceptual interpretation. This inspires our central research question: \textit{Can LLMs be adapted to function not as numerical simulators, but as symbolic analyzers that interpret quantum circuits and articulate their algorithmic logic in an explicit, step-by-step manner?}

In this work, we explore this question by introducing GroverGPT+, an LLM-based framework designed for the symbolic analysis of Grover's algorithm. Here, the term ``GPT'' (\textit{i.e.}, Generative Pre-trained Transformer) is used as a functional shorthand to denote a
transformer-based reasoning model adapted to a specific scientific domain, rather than
implying standard GPT-style pretraining.
 To enable the model to fluently ``read" the language of quantum circuits, we introduce \textbf{quantum-native tokenization}, a method for tokenizing QASM representations by extending the vocabulary of a base tokenizer with quantum-specific operations. To compel the model to ``think step-by-step" and externalize its analysis, we curate a large corpus of training data and employ \textbf{Chain-of-Thought (CoT) supervised fine-tuning}. We choose Grover's algorithm as our primary testbed as its non-trivial, well-defined structure provides an ideal, controlled environment where the accuracy of the model's symbolic reasoning can be rigorously verified against a known analytical solution.

Our results demonstrate that GroverGPT+ can successfully analyze quantum circuits from their QASM representations. Instead of merely outputting final probabilities, the model generates structured, interpretable reasoning traces that correctly identify high-level algorithmic structures, including the oracle and its marked states. This work establishes a benchmark for the task of automated symbolic analysis of quantum circuits and provides empirical evidence of the model's extrapolation behavior. Ultimately, our findings position LLMs as a new class of complementary tools for quantum information science: they are not for replacing numerical simulators, but for aiding in tasks requiring conceptual understanding, such as automated verification, debugging, and education. This work opens a new direction for AI systems that reason about the logic of quantum algorithms, suggesting that an algorithm's ``learnability" can itself provide a new lens for understanding its conceptual complexity, a topic of core interest to quantum information science.


\section{Results and Discussion}

\begin{figure}[htp]
\centering
\includegraphics[width=0.48\textwidth]{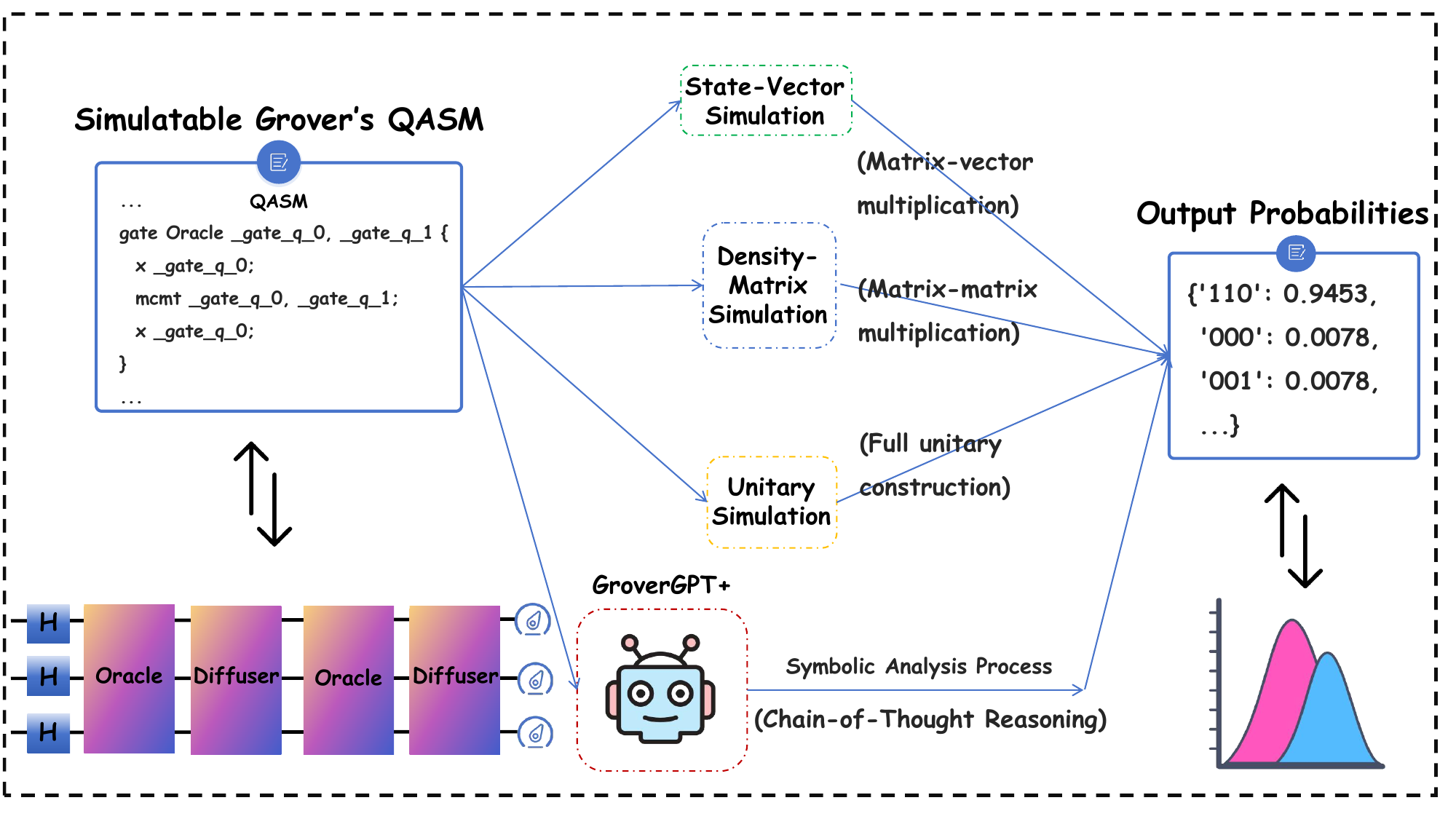}
\vspace{-0.5cm}
\caption{A comparison of two distinct tasks: \textbf{classical simulation} and \textbf{symbolic analysis}. Traditional methods (top path) such as State-Vector simulation  take a QASM input and perform \textbf{numerical operations} like matrix-vector multiplication to directly compute the final output probabilities. In contrast, our approach with GroverGPT+ (bottom path) first performs \textbf{symbolic analysis} using Chain-of-Thought reasoning to generate an \textbf{interpretable reasoning trace} that explains the circuit's logic. The final probabilities are then inferred from this analysis.}
\label{fig:task}
\end{figure}

\begin{figure*}
    \centering
    \includegraphics[width=\textwidth]{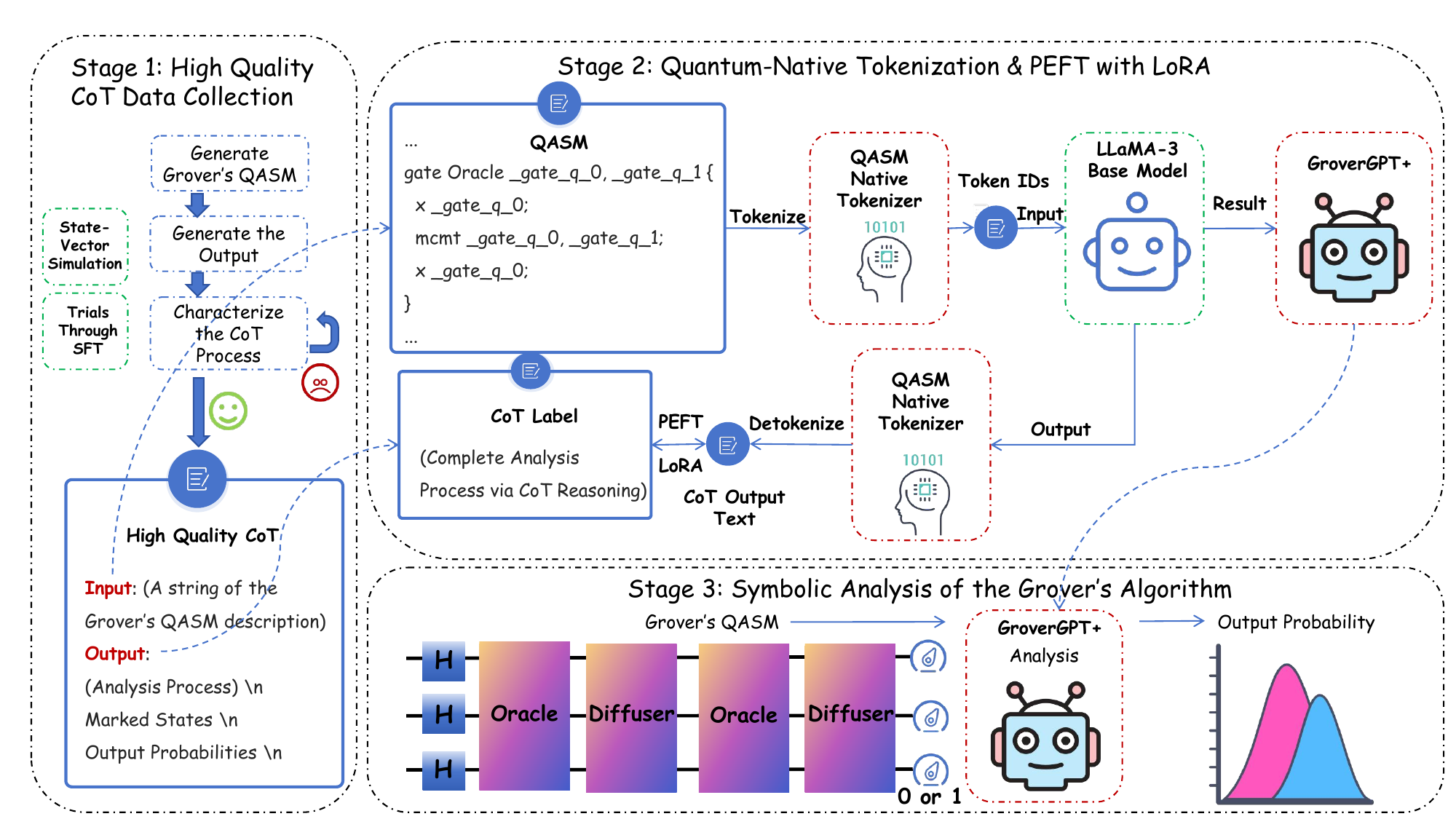}  
    \vspace{-10pt}
    \caption{The overall framework of GroverGPT+ and its application for the symbolic analysis of Grover's algorithm consists of three stages. \underline{Stage 1:} We initiate by collecting high-quality CoT data tailored for Grover’s algorithm. This involves generating Grover’s QASM circuits, performing classical simulations via the state-vector simulation method, and labeling the output distributions along with marked states as CoT supervision targets.
    \underline{Stage 2:} The collected QASM-CoT pairs are tokenized using our QASM-native tokenizer. We then adopt PEFT using the LoRA technique to specialize the base LLM for the symbolic analysis of quantum circuits while maintaining training efficiency.
    \underline{Stage 3:} GroverGPT+ can now serve as a tool for symbolic analysis: given a Grover's QASM circuit, it generates an interpretable reasoning trace that identifies the marked states and infers the final state probability distribution through CoT reasoning.
}
    \label{fig:GroverGPT+}
\end{figure*}

\subsection{Details of the Tasks}

The central task of this work is the symbolic analysis of quantum circuits. Given a circuit representation (\eg, in QASM format), the primary objective is to generate a human-readable reasoning trace that identifies the circuit's high-level algorithmic components. For Grover's algorithm (see \textit{Supplementary Information~1} for a detailed introduction), this corresponds to correctly identifying the oracle and its marked states. This task is distinct from numerical simulation as its principal output is not a final probability distribution, but rather the symbolic, conceptual insight derived from the circuit's structure. A comparison between our symbolic analysis task and traditional classical simulation is illustrated in Figure~\ref{fig:task}. Besides, our model also generates a final probability distribution over the computational basis states. We utilize this numerical output for two purposes: (i) as a scalable metric to quantitatively evaluate the model's ability to identify the marked states, and (ii) as a method for end-to-end validation, where a correct symbolic understanding should lead to a high-fidelity final state distribution. Formally, let $C$ denote a Grover circuit. The goal of symbolic analysis is to produce a reasoning trace $R(C)$ that explicitly identifies the set of marked states, $\mathcal{M}_{\text{true}}$. The predicted probability distribution $g_\theta(C)$ serves to verify the correctness of $R(C)$. Below we introduce the evaluation metrics:

We first introduce Search Accuracy (SA). Given $k = |\mathcal{M}_{\text{true}}|$, we sort candidates by their predicted probabilities $p_i^{\text{model}}$ in descending order, and in case of ties by the integer value of the binary state in ascending order. Let $T_k$ be the top-$k$ states. To avoid spurious hits under near-uniform predictions, we require a minimum confidence threshold $\tau$:

\begin{equation}
    \widehat{\mathcal{M}}_{\text{model}} \;=\; \{\, x \in T_k \;:\; p_x^{\text{model}} \ge \tau \,\}.
\end{equation}

We then define
\begin{equation}
    \text{SA} \;=\; \frac{|\widehat{\mathcal{M}}_{\text{model}} \cap \mathcal{M}_{\text{true}}|}{k}.
\end{equation}

If $|\widehat{\mathcal{M}}_{\text{model}}| < k$, the remaining slots are treated as misses. Notably, the definition above applies generally to arbitrary $m$. In our experimental setting, the number of marked states satisfies $k \leq 3$, we thereby set $\tau = 0.3$ by default. Meanwhile, the models are prompted or trained to output at least the top-30 candidates, so truncation never affects SA, \ie, we have ensured $k \leq t$ denoted as the number of truncated states in practice. If a model outputs fewer candidates, any truncated marked states are treated as misses.

Besides, we introduce Classical Fidelity (CF). Our evaluation focuses on the similarity between \emph{measurement amplitude probability distributions} produced by a method and by an ideal simulator. Given two probability distributions over the computational basis,
$p=(p_1,\dots,p_d)$ and $q=(q_1,\dots,q_d)$ with $d=2^n$, we use the \emph{classical fidelity (CF)}~\cite{nielsen2010quantum,bhattacharyya1943measure,matsumoto2010reverse,starke2024efficient} (see \textit{Supplementary Information~4} for the relation to quantum state fidelity).

\begin{equation}
    CF(p, q)\;=\;\Big(\sum_{i=1}^{d}\sqrt{p_i q_i}\Big)^2.
\label{eq:CF}
\end{equation}

When a model outputs a truncated distribution (e.g., top-30 states for baseline LLMs), we treat any state not present in the model's output as having zero probability. Specifically, for the model's predicted distribution $p$, if a computational basis state $i$ is not included in the output, we set $p_i = 0$. The CF is then computed using Equation~(\ref{eq:CF}) over all $d = 2^n$ states, where missing states contribute zero to the sum. This approach ensures that truncation does not artificially inflate fidelity scores, as states omitted from the output are penalized through their zero probability assignment.

\subsection{Overview of GroverGPT+}

Figure~\ref{fig:GroverGPT+} presents the overall framework of GroverGPT+. It is an LLM with 8 billion parameters supervised fine-tuned on the base Llama-3~\citep{grattafiori2024llama} model. To conduct our task, we firstly develop GroverGPT+ through stages including high-quality CoT data collection (\underline{Stage 1}), quantum-native tokenization and parameter-efficient fine-tuning (PEFT) with low-rank adaptation (LoRA) (\underline{Stage 2}), and then perform symbolic analysis of Grover's algorithm (\underline{Stage 3}). Below are the details for each stage:

In \underline{Stage 1}, we first generate high-quality CoT training data. Grover's QASM circuits are generated starting from 2 qubits, marking 1 to 3 target states. For each circuit size, the number of marked states never exceeds the number of qubits and is capped at three. See \textit{Supplementary Information~12} for the detailed experimental setup. Corresponding probability amplitudes are computed using brute-force state-vector simulation. CoT processes are then annotated based on outputs from an intermediate supervised fine-tuned LLM. We finalize the curation of the dataset once desirable CoT processes are observed.

In \underline{Stage 2}, we supervised fine-tune GroverGPT+ using PEFT with LoRA. Initially, collected QASM descriptions are tokenized into token IDs using our quantum-native tokenizer (detailed in Section~\ref{sec:method:qasm} and \textit{Supplementary Information~6}). These token IDs serve as inputs to the LLaMA-3 base model, whose outputs are then detokenized into a text format. PEFT with LoRA is conducted for higher training efficiency. 

In \underline{Stage 3}, once trained, GroverGPT+ accepts Grover’s QASM descriptions as input and performs symbolic analysis via CoT reasoning. The model outputs structured text including intermediate reasoning steps, marked states, and the output probability amplitudes of all computational basis states. Specifically, the complete CoT process is detailed in \textit{Supplementary Information 8}.

When analyzing Grover's algorithm, GroverGPT+ only requires a pure QASM description of a quantum circuit as input, without additional information, while general-purpose LLMs need a meticulous prompt design to guide the LLM to output correct results. Therefore, GroverGPT+ offers a more streamlined and efficient workflow for this analysis task. Below briefly introduces how this is achieved:

Firstly, GroverGPT+ extracts the Oracle entity from the whole bunch of long QASM for searching the marked computational basis states in the following steps. Secondly, GroverGPT+ reasons about each corresponding marked state according to the oracle construction extracted before. GroverGPT+ leverages how the target states are marked according to Grover's algorithm design. Thirdly, following the second step, GroverGPT+ outputs the probabilities of the marked states and the unmarked states according to the reasoned information. It is achieved through a learned mapping from basic information including the number of qubits, the number of marked states, and the searched results from the previous steps, to the probability amplitudes for each computational basis state.

\subsection{Experimental Settings}
\label{Experimental Results}

We first introduce the general experiment settings, then we conduct empirical studies of GroverGPT+ with respect to its in-distribution and nearby out-of-distribution (OOD) performance. Furthermore, we also evaluate its computational scalability. Notably, evaluations with respect to its CoT advantage, the technique of quantum-native tokenization and its extrapolation performance are respectively shown in \textit{Supplementary Information 13}, \textit{Supplementary Information 14} and \textit{Supplementary Information 15}. The hyperparameter settings for this section can be found in \textit{Supplementary Information 16}. 

We first detail the general experimental settings. We evaluate GroverGPT+ under different input formats and quantum circuit settings. Specifically, we consider two types of inputs: \textbf{Full-circuit Input} and \textbf{Oracle-only Input}, each designed to probe different aspects of model capability. The experiments vary the number of qubits and marked states to comprehensively assess performance, while also exploring the model's scaling behavior under increasing circuit sizes. For detailed configurations, see \textit{Supplementary Information~12}. Notably, for the oracle-only input setting, the target distribution is defined using the analytically optimal iteration number $k_{\text{opt}}$, which guarantees the maximum success probability for a given $(n, t)$ configuration, where $n$ is the number of qubits and $t$ is the number of marked states. This ensures that the output distributions are consistently defined even when the iteration number $k$ is not explicitly present in the oracle-only input. In all plots, solid lines denote the mean over runs, and discrete error bars indicate mean~$\pm$~one standard deviation~(std) for both SA and CF.

\subsection{Empirical Study of GroverGPT+ in Analyzing Grover's Algorithm}
\label{sec:simulation}

\begin{figure}[h!]
\centering
\includegraphics[width=0.48\textwidth]{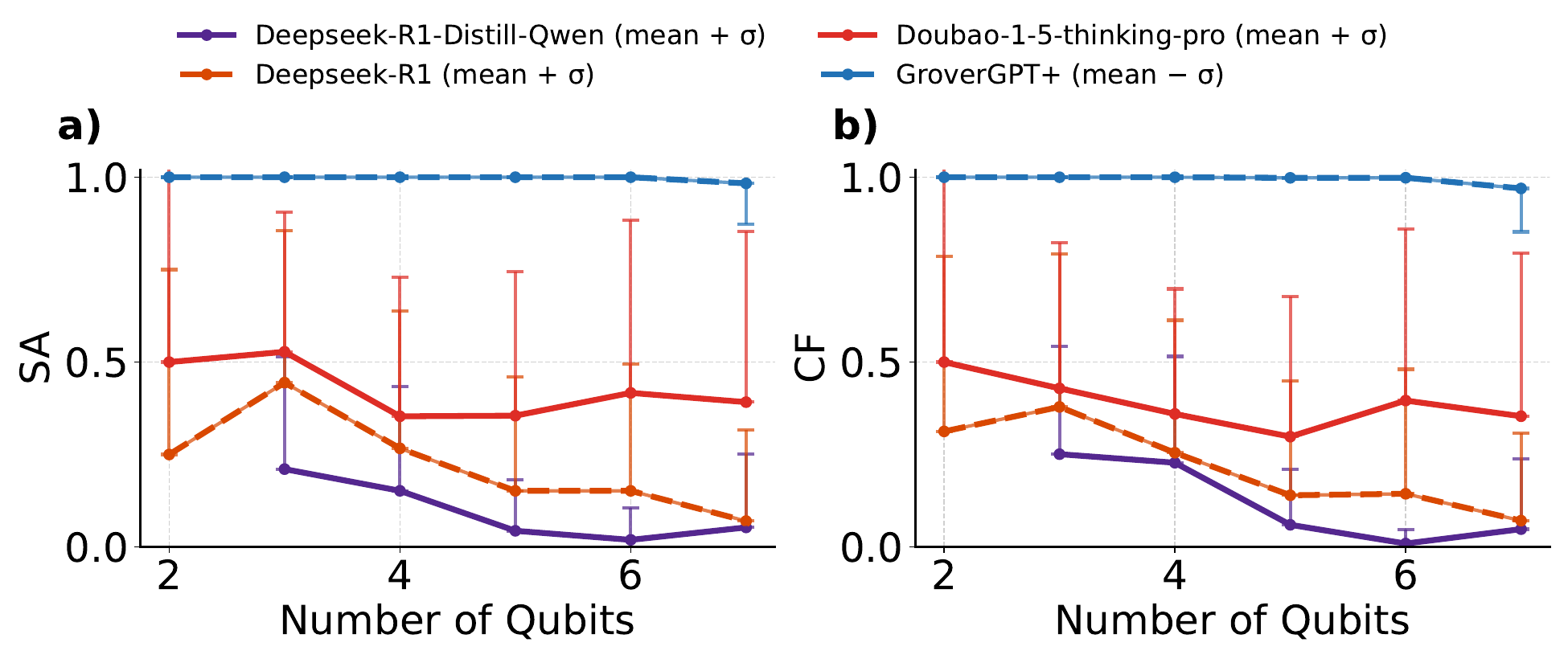}
\vspace{-0.5cm}
\caption{Performance of GroverGPT+ against baseline LLMs in the symbolic analysis of Grover's algorithm in terms of (a) SA and (b) CF, across varying numbers of qubits. Solid lines show means; discrete error bars indicate uncertainty (mean $+$ std (\(\sigma\)) or, where noted, mean$-$\,std).}
\label{fig:transductive}
\end{figure}

\begin{figure}[h!]
\centering
\includegraphics[width=0.48\textwidth]{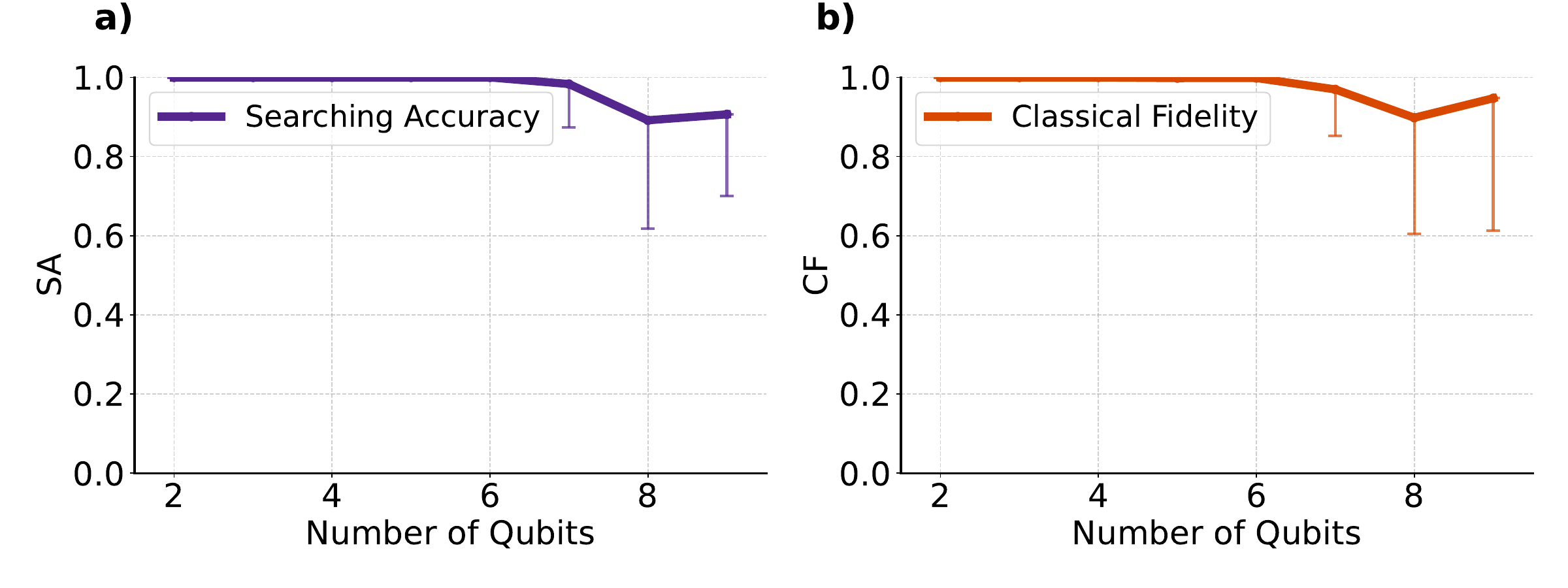}
\vspace{-0.5cm}
\caption{Nearby OOD performance of GroverGPT+ when scaling up to 8 and 9 qubits (beyond the training range). Both the SA (a) and CF (b) serve as the evaluation metrics. Solid lines show means; discrete error bars indicate uncertainty (mean$-$\,std).} 
\label{fig:generalization}
\end{figure}

\begin{figure}[h!]
\centering
\includegraphics[width=0.48\textwidth]{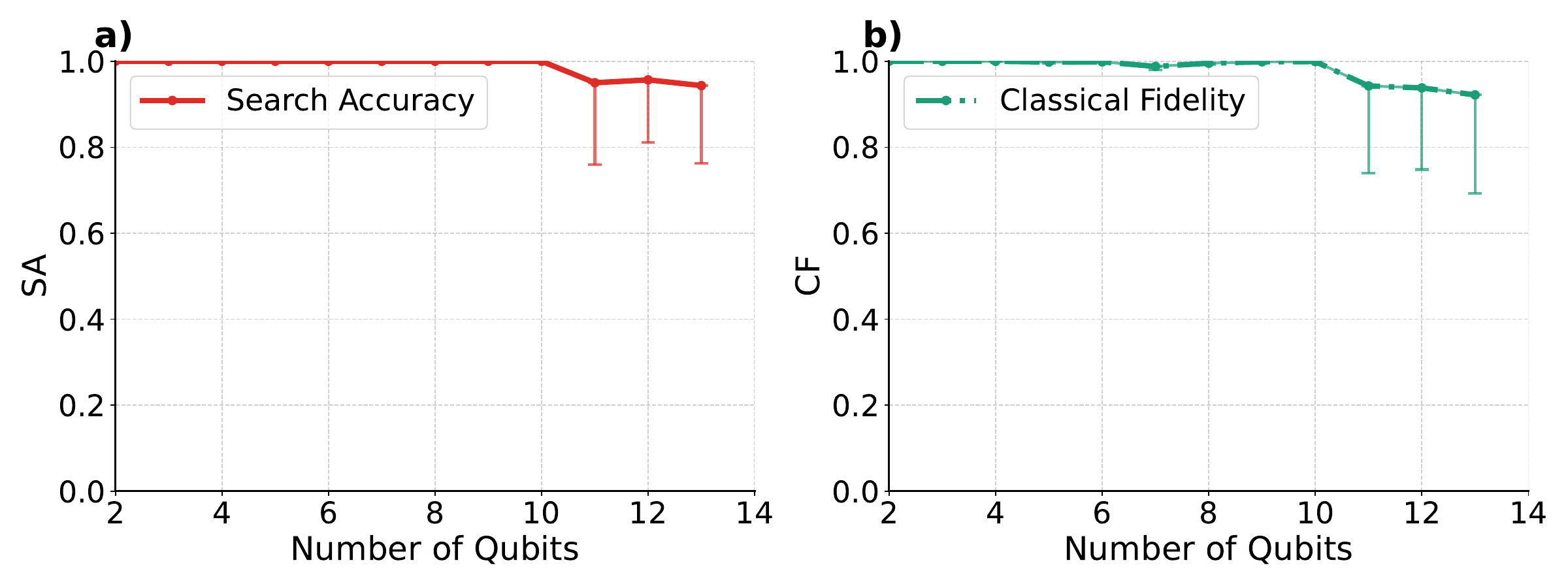}
\vspace{-0.5cm}
\caption{The performance of GroverGPT+ under the Oracle-only input setting with the number of qubits \(n = \{2, 3, ..., 13\}\). Both the SA (a) and the CF (b) serve as the evaluation metrics. Solid lines show means; discrete error bars indicate uncertainty (mean$-$\,std).}
\label{fig:oracle_only_input_main}
\end{figure}


In this section, we empirically evaluate GroverGPT+'s performance in the symbolic analysis of Grover's quantum search algorithm. We first test its ability using \textbf{full-circuit inputs} for qubit counts \(n \in \{2, 3, \dots, 7\}\), corresponding to the training data. Given that the maximum token length is exceeded at \(n = 9\) (see \textit{Supplementary Information 15}), we separately assess GroverGPT+'s generalization performance at $n = \{8, 9\}$ and compare it with the trained scenarios. Additionally, we evaluate the performance on analyzing \textbf{oracle-only inputs} across a broader range, \(n \in \{2, 3, \dots, 13\}\). The results are respectively shown in Figure~\ref{fig:transductive}, Figure~\ref{fig:generalization} and Figure~\ref{fig:oracle_only_input_main}. Below are observations regarding the results:

Figure~\ref{fig:transductive} illustrates that baseline LLMs exhibit relatively low SA and fidelity values (around 0.2–0.5) with substantial standard deviations, particularly for qubit counts between 5 and 7. For instance, at 7 qubits, baseline models achieve SA and fidelity below 0.4, indicating unstable performance. Conversely, GroverGPT+ consistently attains high SA and fidelity values approaching 1.0 with minimal variability, highlighting its stability and superior performance. This advantage likely results from GroverGPT+'s specialized fine-tuning using high-quality CoT data and quantum-native tokenization, enhancing result consistency across varying circuit sizes.

Figure~\ref{fig:generalization} evaluates GroverGPT+ on qubit counts slightly beyond the training range by training the model up to 7 qubits and test it on 8–9 qubits, \ie, a nearby OOD by qubit count setting. We can observe a mild drop in SA $\approx 0.89$ with 8 qubits and SA $\approx 0.91$ with 9 qubits, while the CF remains above $0.90$. These results indicate that the model is not merely memorizing the training distribution and exhibits potential to scale to larger circuits within the tested range (2--9 qubits).

Figure~\ref{fig:oracle_only_input_main} further examines the Oracle-only input format on 10–13 qubits, where both SA and the CF remain close to $1.0$. We view this as encouraging evidence that the compact Oracle-only representation supports evaluation at larger qubit counts under context-length constraints within the tested range (2--13 qubits), indicating potential scalability within this tested regime.

Notably, the evaluations are based on both single-target and multi-target Grover circuits. Compared to the single-target case, multi-target Grover’s algorithm requires multiple oracle blocks, each responsible for flipping the phase of one marked state. This leads to longer and structurally more complex QASM circuits. Consequently, the CoT reasoning produced by the model must also handle multiple oracle blocks, increasing the length and potential variability of reasoning chains. This may in turn reduce both SA by missing one or more marked states and CF by spreading probability mass over incorrect states. While single- and multi-target Grover circuits differ in their QASM representations, we emphasize that our evaluation does not hinge on this distinction. We are tackling an analysis scenario where the inputs are direct simulatable QASMs, thereby the number of marked states is not known \emph{a priori}, and hence the task setting naturally mixes single- and multi-target instances, which makes the evaluation closer to realistic scenarios. Accordingly, when reporting the SA and CF, we aggregate results by computing the mean and standard deviation across both single- and multi-target instances rather than isolating them.

\subsection{Computational Scalability of GroverGPT+}
\label{sec:Computational_Scalability}

\begin{figure}[h!]
\centering
\includegraphics[width=0.48\textwidth]{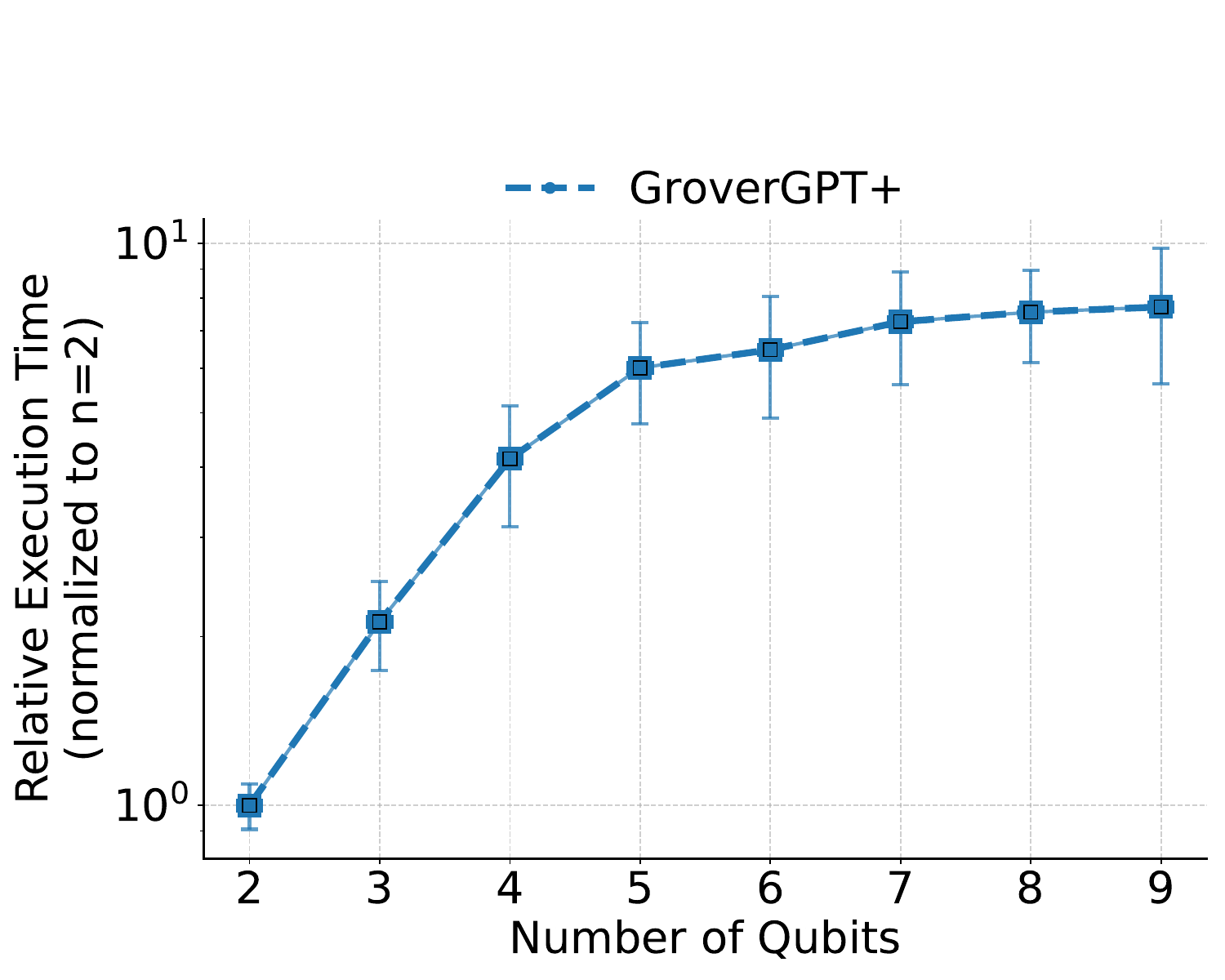} 
\vspace{-0.5cm}
\caption{Computational scalability of GroverGPT+. The plot shows the relative inference time for the symbolic analysis task, normalized to the runtime at 2 qubits, as a function of the number of qubits. Solid lines show the mean over three runs; discrete error bars indicate mean $\pm$\, one standard deviation.}
\label{fig:computational_scalability} 
\end{figure}

In this section, we characterize the \emph{computational scalability} of the symbolic analysis task performed by GroverGPT+. We define this as the growth of the model's inference time with respect to the number of qubits, $n$. To highlight the scaling trend independently of hardware-specific overheads, we measure the relative execution time. Formally, let $T(n)$ denote the mean inference time for an $n$-qubit circuit. We define the scalability metric as:
\begin{equation}
    S(n) \;=\; \frac{T(n)}{T(2)},
\end{equation}
which normalizes the runtime to its value at $n=2$. The absolute execution times are also evaluated in \textit{Supplementary Information 7} for completeness.

The experiments were conducted by measuring the inference time of GroverGPT+ across qubit sizes ranging from 2 to 9. For each qubit count, three runs were performed on an NVIDIA RTX A6000 GPU with 48 GB GDDR6 VRAM to compute the mean and standard deviation. The reported runtimes refer to inference time only and do not include the one-time cost of model training. Refer to \textit{Supplementary Information~7} for more details.

The results are plotted on a logarithmic scale in Figure~\ref{fig:computational_scalability}. Notably, all scaling observations reported in this section are limited to the tested range. We observe that the execution time of GroverGPT+ exhibits a gentle and notably sub-linear growth trend with respect to the number of qubits. The relative execution time remains consistently within a 1--10$\times$ range compared to its baseline at 2 qubits, and the variance remains stable. This favorable scaling is a direct consequence of our symbolic analysis approach. Instead of performing tensor-based state evolution, which is inherently exponential, GroverGPT+'s CoT reasoning operates on the symbolic structure of the QASM input. This allows the model to avoid the exponential computational overhead associated with methods that must track the full $2^n$-dimensional state vector, \ie, GroverGPT+ can also serve as a useful tool for finding the final output probability given the simulatable QASM within the tested regimes. To further explore, since GroverGPT+ directly operates on simulatable QASM inputs and finally generates the output amplitude distributions, we also compare this end-to-end latency with some traditional classical simulations. Refer to \textit{Supplementary Information 7} for details.

\subsection{Discussion}

Our work began with a central research question: \textit{Can LLMs be adapted to function not as numerical simulators, but as symbolic analyzers that interpret quantum circuits and articulate their algorithmic logic in an explicit, step-by-step manner?} Our findings provide an answer for this. Through the development of GroverGPT+, we have demonstrated that an LLM, when equipped with domain-specific techniques like quantum-native tokenization and structured Chain-of-Thought training, can successfully parse low-level QASM code and produce high-level, interpretable reasoning traces. These traces are not merely a byproduct; they are the primary output, revealing the model's "understanding" of algorithmic components like the oracle and its marked states. Our work thus is complementary to  prior studies such as GroverGPT~\citep{wang2024grovergpt} (see \textit{Supplementary Information 3} for detailed comparison) by shifting the objective from reproducing numerical outcomes to elucidating the underlying logical process, establishing a stringent benchmark for the automated analysis of quantum algorithms.


For the domain of quantum information science, our study points towards more than a practical tool: it suggests a promising avenue for future research in evaluating algorithmic complexity through a new conceptual lens. While traditional metrics focus on physical resources like gate count and circuit depth, a compelling future direction would be to investigate whether the `learnability' of an algorithm by a general-purpose reasoner like an LLM can serve as a proxy for its descriptive or conceptual complexity. Our findings on computational scalability provide initial evidence for this direction. We observed that the inference time for analyzing Grover's algorithm known for its highly regular and iterative logic scales sub-linearly with the number of qubits. This favorable scaling indicates the model's effort might be tied to the algorithm's low conceptual complexity, not the exponential size of its Hilbert space. This framework opens new research questions for quantum information theory: would analyzing a complex Variational Quantum Eigensolver (VQE) ansatz or identifying stabilizer generators in a quantum error correction (QEC) code reveal a higher conceptual complexity through this new lens? This approach begins to reframe the LLM from a simple tool into a scientific instrument for probing the nature of quantum algorithms themselves, marking an early and encouraging exploration into a new, AI-driven approach to theoretical quantum information science.

Besides, our work contributes a generalizable methodology for applying LLMs to the symbolic analysis of quantum circuits. For example, our \textbf{quantum-native tokenization} is not specific to Grover's algorithm. Instead it can be adopted to efficiently represent any QASM-described circuit in a way that is semantically meaningful to a transformer architecture. Similarly, our strategy for \textbf{structure-aware CoT fine-tuning} serves as a template for teaching an LLM the specific logical steps of other quantum algorithms (see \textit{Supplementary Information 8}). Together, these techniques form a foundational toolkit that enables future research into the automated analysis of a much wider range of quantum algorithms, from the Quantum Fourier Transform to complex variational circuits, thereby paving the way for more sophisticated AI-driven tools in the quantum domain.


\section{Methods}

\subsection{Quantum-Native Tokenization}\label{sec:method:qasm}

\begin{figure}[h!]
\centering
\includegraphics[width=0.48\textwidth]{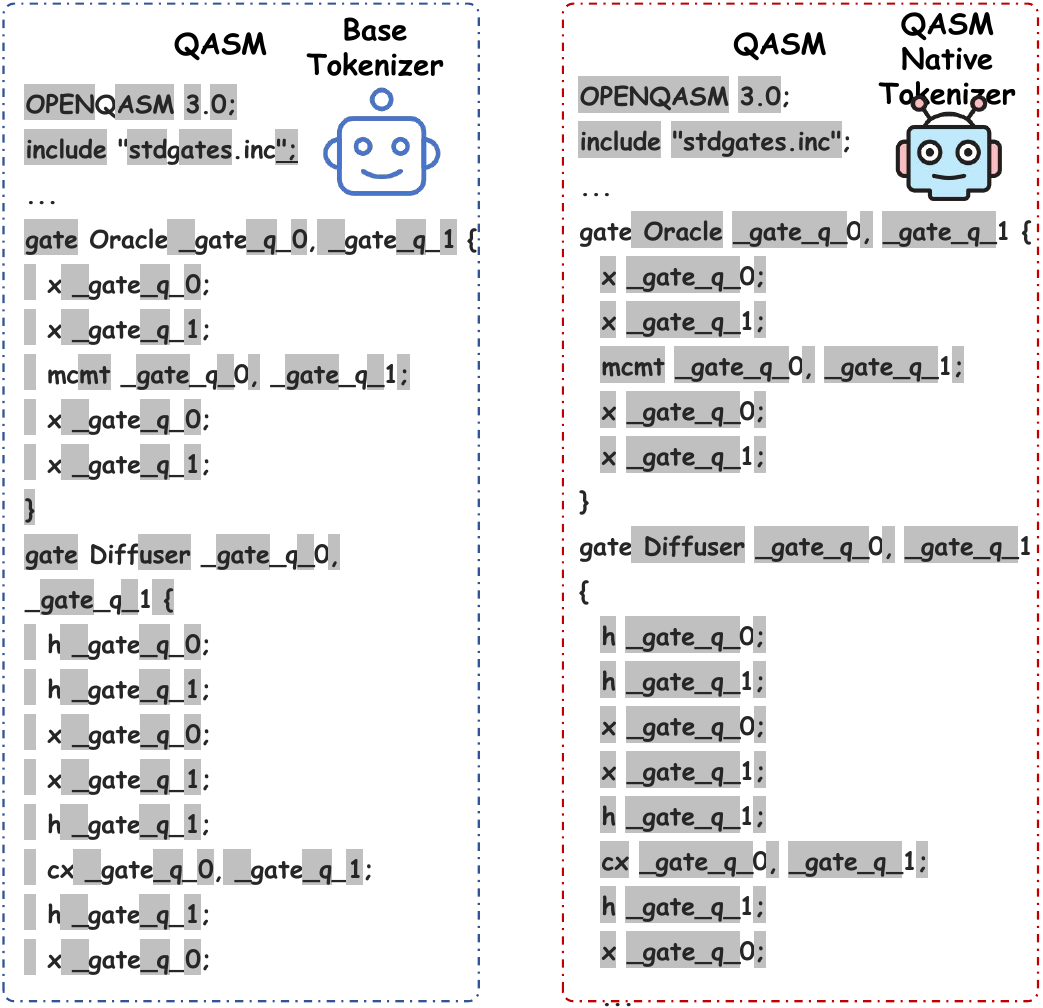}
\vspace{-0.5cm}
\caption{Comparison of the base tokenizer and the quantum-native tokenizer on QASM input. Each grey and non-grey segment represents a distinct token. The base tokenizer fragments the syntax into subword units, while the quantum-native tokenizer preserves gate operations and qubit references as cohesive tokens, resulting in more compact and efficient representations.}
\label{fig:base_quantum_tokenizer_compare}
\end{figure}

The LLaMA model is primarily trained on approximately 1.4T English-language tokens sourced from the Internet~\citep{touvron2023llama}, lacking native support for quantum-specific languages such as QASM. As a result, it struggles to tokenize QASM code effectively, leading to fragmented subword sequences that ignore the language's syntactic and semantic structure. This inefficient tokenization increases input length and memory usage. As illustrated in Figure~\ref{fig:base_quantum_tokenizer_compare}, the base tokenizer breaks down QASM statements into disjointed pieces based on natural language rules, rather than recognizing them as coherent units. To overcome these limitations, we propose a quantum-native tokenizer tailored to the structure of quantum programming languages. Specifically designed for QASM, this tokenizer captures key elements—such as gate operations, qubit identifiers, and block constructs—as discrete, semantically meaningful tokens. By aligning with the intrinsic structure of QASM, it achieves more compact and efficient tokenization, reducing context length and improving memory efficiency in downstream tasks. The development process is detailed as follows:

Firstly, we collect a large-scale dataset encompassing a comprehensive range of QASM circuit descriptions, covering quantum circuits with qubit numbers ranging from \(n = 2\) to \(n = 9\). Secondly, to systematically process and analyze these QASM circuits, we develop a set of custom parsing rules tailored to the unique syntactic structure of QASM. These rules tokenize each line of the QASM files to accurately extract quantum gate definitions and operation commands. Specifically, our rule-based approach first identifies gate definitions using regular expressions that capture gate names, parameters, and qubit arguments; any numerical suffixes specific to certain internal naming conventions (\eg, \textit{\_gate\_q\_}, \textit{unitary\_}, \textit{mcx\_vchain\_}) are stripped to maintain consistency and generality in subsequent analyses. For standard quantum commands, we parse the operation names, optional parameters, and target qubits separately. These components are then tokenized, again removing extraneous numerical suffixes to ensure uniformity. The parsing mechanism also explicitly handles structural delimiters, such as opening and closing braces, crucial for correctly interpreting nested gate definitions and circuit hierarchies.

These custom rules enable scalable and automated preprocessing of QASM descriptions, facilitating efficient symbolic analysis. Figure~\ref{fig:base_quantum_tokenizer_compare} also presents an example of how a single QASM description is tokenized using the base tokenizer and the quantum-native tokenizer, which pinpoints the brought efficiency. In total, we extend 266 specific vocabularies that contain complete semantics, such as \textit{mcx} indicating a multi-controlled X gate. The rule definitions, together with their corresponding Python implementations, are elaborated in \textit{Supplementary Information~6}.

\subsection{Chain-of-Thought Training}\label{sec:method-cot}

Chain-of-Thought (CoT) reasoning is an emergent capability in LLMs (see \textit{Supplementary Information~2} for introductions of LLM-related techniques), allowing them to solve complex problems through step-by-step deduction~\citep{wei2022emergentabilitieslargelanguage,wei2023chainofthoughtpromptingelicitsreasoning}. Formally, given a prompt $\mathbf{Q}$, CoT augments the output by generating intermediate reasoning steps $\{c_1, c_2, \ldots, c_n\}$ before producing the final answer $\mathbf{A}$:
\begin{align}
\text{CoT}(\mathbf{Q}) = \{[c_1, c_2, \ldots, c_n], \mathbf{A}\}.
\end{align}

To improve GroverGPT+'s reasoning in the symbolic analysis of quantum circuits, we adopt explicit CoT training, where intermediate reasoning chains are included in the supervision training targets. Unlike prior work~\citep{wang2024grovergpt} that directly predicts output probabilities, our approach models the full analysis process as a sequence of logical deductions.

We construct two types of CoT training datasets. The first, CoT Data with Oracle-only Input, includes only oracle QASM code as input, with CoT reasoning sequences as targets; this design encourages GroverGPT+ to reason directly from oracle structure. The second, CoT Data with Full-circuit Input, includes the full Grover circuit QASM code as input, paired with CoT outputs to enhance accuracy and focus on oracle extraction. These datasets span various qubit ranges to ensure broad generalization. We detail our CoT training technique in \textit{Supplementary Information~9}. \\

\section*{Acknowledgments}
MC, JC, and JL are supported in part by the University of Pittsburgh, School of Computing and Information, Department of Computer Science, Pitt Cyber, PQI Community Collaboration Awards, Pitt Momentum fund, and John C. Mascaro Faculty Scholar in Sustainability. This research used resources of the Oak Ridge Leadership Computing Facility, which is a DOE Office of Science User Facility supported under Contract DE-AC05-00OR22725. PL and TC are partially supported by Amazon Research Award, UNC Accelerating AI Awards, NAIRR Pilot Award, OpenAI Researcher Access Award, and Gemma Academic Program GCP Credit Award. \\

\section*{Data Availability Statement}

The data supporting the findings of this study are available at \url{https://github.com/mchen644/GroverGPT-plus}.

\section*{Code Availability Statement}

The results are reproducible with the code available at \url{https://github.com/mchen644/GroverGPT-plus}.

\section*{Competing Interests}

J.L. is an associate editor of npj Quantum Information, but were not involved in the editorial review of, or the decision to publish this article. All other authors declare no competing interests.

\section*{Author Contributions}

J.L and T.C proposed the study of large language model in symbolic analysis of Grover's algorithm. M.C proposed the methods of quantum-native tokenization and adopted Chain-of-Thought technique. M.C. designed and performed the experiments, with support from J.C., P.L. and H.W., under the supervision of J.L. and T.C.. M.C., J.C and P.L wrote the manuscript, with inputs and contributions from all authors.

\clearpage
\onecolumngrid

\renewcommand{\figurename}{Supplementary Figure}
\renewcommand{\tablename}{Supplementary Table}
\renewcommand{\lstlistingname}{Supplementary Note}

\appendix
\setcounter{section}{0}
\renewcommand\thesection{\arabic{section}}

\setcounter{secnumdepth}{3}
\setcounter{tocdepth}{2}

\section*{Supplementary Information}

\section{Grover's Quantum Searching Algorithm}
\label{appendix:Grover's_Quantum_Searching_Algorithm}

Grover's quantum search algorithm theoretically provides a quadratic speed-up over its classical counterpart for unstructured search problems. Specifically, classical searching requires \(\mathcal{O}(N)\) query complexity to address a computational problem with the input that can be accessed only through the queries, while Grover's algorithm only uses \(\mathcal{O}(\sqrt{N})\) evaluations. This advantage is significant when there is a big problem size \(N\).

Generally, there are two main steps to construct the Grover's quantum circuits: the Oracle construction and the diffuser operation construction. The Oracle is responsible for applying a selective phase flip to a designated marked state in the computational basis, which is a critical step in Grover's algorithm. Our implementation of the oracle construction follows the design principles introduced in the QCircuitNet framework~\citep{yang2024qcircuitnet}. The Python implementation is shown in Supplementary Note \ref{code:oracle_construction}. Specifically, given a bit-string representing the marked state, the bit-string is first reversed to match the internal qubit indexing convention adopted in Qiskit. This adjustment ensures that the circuit operations are correctly aligned with the intended computational basis states. Next, we identify the qubits corresponding to the '0' entries in the reversed bit-string. An \(X\) gate is applied to each of these qubits, thereby transforming the marked state into the all-ones state \(|11\ldots1\rangle\). Following this transformation, we apply a multi-controlled \(Z\) gate (MCMT-\(Z\)) to the circuit. This gate flips the phase of the \(|11\ldots1\rangle\) state while leaving all other basis states unchanged. The controlled operation ensures that only the transformed marked state acquires a phase of \(-1\). After the phase flip, the same \(X\) gates are reapplied to the qubits to revert the computational basis back to its original configuration. Finally, the resulting circuit is converted into a quantum gate object named "Oracle," which can be seamlessly integrated into the broader Grover search framework. This construction guarantees that the oracle satisfies the necessary condition of Grover’s algorithm: applying a \(-1\) phase to the marked state while leaving the others invariant. 

\begin{lstlisting}[style=quantumcode, language=Python, caption={Python implementation for the oracle construction under single marked state}, label=code:oracle_construction]
def create_oracle(n, marked_state):
    oracle = QuantumCircuit(n)
    rev_target = marked_state[::-1]
    zero_inds = [ind for ind, char in enumerate(rev_target) if char == "0"]
    if zero_inds:
        oracle.x(zero_inds)
    oracle.compose(MCMT(ZGate(), n - 1, 1), inplace=True)
    if zero_inds:
        oracle.x(zero_inds)
    oracle_gate = oracle.to_gate()
    oracle_gate.name = "Oracle"
    return oracle_gate
\end{lstlisting}





Here, we extend the implementation to the case of multiple marked states in Supplementary Note~\ref{code:oracle_construction_MMS}.  
Instead of constructing a single oracle for one marked state, we iterate over a set of marked states and apply the corresponding oracle subroutine for each. For each bit-string in the list of marked states, we follow the same transformation as in the single-state case: the bit-string is first reversed to match Qiskit's internal qubit ordering, and \(X\) gates are applied to the qubits corresponding to '0' entries to convert open controls into closed controls.  
Afterward, a multi-controlled \(Z\) gate is applied to flip the phase of the mapped \(|11\ldots1\rangle\) state.  
The applied \(X\) gates are then reversed to restore the original computational basis.  
By repeating this process across all marked states, the resulting oracle circuit introduces a \(-1\) phase to each of the target states, allowing Grover’s algorithm to function in the multiple marked states searching setting.

\begin{lstlisting}[style=quantumcode, language=Python, caption={Python implementation for the oracle construction under multiple marked states}, label=code:oracle_construction_MMS]
def create_oracle(n, marked_states):
    """Create an oracle circuit for multiple marked states."""
    
    # Initialize an empty oracle circuit with n qubits
    oracle = QuantumCircuit(n)
    
    # Loop through each marked state to construct corresponding oracle operations
    for marked_state in marked_states:
        # Reverse the marked state bit-string to match Qiskit's bit-ordering convention
        rev_target = marked_state[::-1]
        
        # Find indices of bits set to '0' (these require open controls)
        zero_inds = [ind for ind, char in enumerate(rev_target) if char == "0"]
        
        # Apply X-gates on qubits corresponding to '0' to convert open controls to closed controls
        if zero_inds:
            oracle.x(zero_inds)
        
        # Apply multi-controlled Z gate (MCMT) with n-1 controls and 1 target qubit
        oracle.compose(MCMT(ZGate(), n - 1, 1), inplace=True)
        
        # Re-apply X-gates to restore qubits to original state
        if zero_inds:
            oracle.x(zero_inds)
    
    # Convert constructed oracle to a gate object for modular use
    oracle_gate = oracle.to_gate()
    oracle_gate.name = "Oracle"
    
    return oracle_gate

\end{lstlisting}


\section{Large Language Models, Their Supervised Fine-Tuning and Parameter-Efficient Fine-Tuning}
\label{appendix:Large Language Models, Their Supervised Fine-Tuning and Parameter-Efficient Fine-Tuning}

\textbf{General-purpose Large Language Models~(LLMs).} LLMs have emerged as powerful computational systems capable of understanding and generating human language at a giant scale. 
These models, built upon the Transformer architecture~\citep{vaswani2023attentionneed}, leverage self-attention mechanisms to capture complex linguistic patterns and semantic relationships. 
Modern LLMs such as LLaMA~\citep{grattafiori2024llama3herdmodels}, GPT~\citep{openai2024gpt4technicalreport}, and DeepSeek~\citep{deepseekai2025deepseekr1incentivizingreasoningcapability}, contain billions of parameters trained on vast textual corpora using self-supervised learning objectives. 
The auto-regressive pre-training approach establishes their general linguistic capabilities through next-token prediction tasks, creating representations that capture syntactic structures, factual knowledge, and even reasoning capabilities~\citep{nachane2024shotchainofthoughtdrivenreasoning, wei2022emergentabilitieslargelanguage, deepseekai2025deepseekr1incentivizingreasoningcapability}. 
This architecture enables LLMs to generalize across diverse domains, making them suitable for adaptation to specialized applications such as quantum computing, where our work demonstrates that LLMs can effectively learn and reason about quantum algorithms~(\textit{e.g.}. Grover's search) when provided appropriate QASM code examples and chain-of-thought demonstrations.

\textbf{LLMs for Quantum Computing.} Recently, an increasing number of studies have begun exploring how LLMs can contribute to the field of quantum computing. QGAS~\citep{liang2023unleashing} proposes high-performance ansatz architectures tailored for quantum chemistry and quantum finance tasks. GPT-QE~\citep{nakaji2024generative} focuses on generating quantum circuits with specific desired properties for quantum simulations. AdaInit~\citep{zhuang2025large} alleviates the barren plateau problem~\citep{cerezo2022challenges, cerezo2021cost, arrasmith2021effect, liu2024laziness} in quantum machine learning by providing effective initialization parameters for quantum neural network models. QAOA-GPT~\citep{Tyagin2025uwm} demonstrates the potential of the Generative Pre-trained Transformer framework to generate high-quality quantum circuits for solving quadratic unconstrained binary optimization (QUBO) problems. GroverGPT~\citep{wang2024grovergpt} explores the boundary of classical simulatability by leveraging the pattern recognition capabilities of LLMs and novel prompt design strategies to simulate Grover's algorithm. Building upon this, GroverGPT+ eliminates the need for explicit prompt guidance, introducing techniques that reveal how LLMs can comprehend the underlying logic of quantum algorithms while effectively simulating Grover’s algorithm.

\textbf{Supervised Fine-Tuning~(SFT).} SFT refines pre-trained LLMs for specific tasks using high-quality labeled data. 
Unlike pre-training, which relies on self-supervised objectives, SFT applies direct supervision with input-output pairs curated to guide model behavior toward desired outputs~\cite{hu2021loralowrankadaptationlarge, li2024glidergloballocalinstructiondriven, raffel2023exploringlimitstransferlearning}. 
This process typically requires significantly fewer samples than pre-training but depends critically on data quality and alignment with target applications. 
SFT has proven effective for adapting general LLMs to specialized domains, including programming~\citep{rozière2024codellamaopenfoundation,deepseekai2025deepseekv3technicalreport,kevian2024capabilitieslargelanguagemodels,zhao2024modelgluedemocratizedllmscaling}, mathematics~\citep{kevian2024capabilitieslargelanguagemodels,zhao2024modelgluedemocratizedllmscaling,openai2024gpt4technicalreport}, and scientific applications~\cite{wang2023huatuotuningllamamodel,yun2024flexmoemodelingarbitrarymodality}. 
The technique updates the model's weights to better align with domain-specific knowledge while preserving general capabilities established during pre-training. 
In our work, we leverage SFT to adapt LLaMA to understand quantum computing patterns in QASM format and analyze circuits and infer outputs of Grover's algorithm.

\textbf{Parameter-Efficient Fine-Tuning~(PEFT).} PEFT addresses computational limitations of conventional full LLM model fine-tuning by updating only a small subset of the model parameters while keeping most weights frozen~\citep{hu2021loralowrankadaptationlarge,li2024glidergloballocalinstructiondriven,liu2022ptuningv2prompttuning,li2021prefixtuningoptimizingcontinuousprompts}. 
Techniques such as Low-Rank Adaptation~(LoRA)~\citep{hu2021loralowrankadaptationlarge}, Prefix Tuning~\citep{li2021prefixtuningoptimizingcontinuousprompts}, and adapter modules~\citep{houlsby2019parameterefficienttransferlearningnlp} significantly reduce memory requirements and computational costs while maintaining performance comparable to full fine-tuning. 
PEFT methods typically introduce trainable matrices that modify the forward pass of frozen layers through low-rank decomposition or adapter architectures. 
These approaches have democratized LLM adaptation by enabling fine-tuning on consumer hardware and facilitating efficient domain adaptation~\citep{dettmers2023qloraefficientfinetuningquantized}. 
Our work leverages LoRA to adapt LLaMA to the quantum computing domain~(\textit{i.e.} Grover's algorithm) efficiently, enabling the model to understand QASM representations of Grover's algorithm, generate appropriate chain-of-thought reasoning, and generate outputs through symbolic analysis.



\section{Comparison Between GroverGPT and GroverGPT+}
\label{appendix:grovergpt_comparison}


We provide a comparison between GroverGPT~\cite{wang2024grovergpt} and our proposed GroverGPT+. Supplementary Table~\ref{tab:grovergpt_comparison} summarizes the key distinctions across task definition, evaluation focus, model scale, data scale, and the relative parameter increase induced by tokenizer extension. This highlights that the two methods should be regarded as complementary. Regarding the extra parameters, we estimate through following calculations:

LLaMA-8B has a hidden size or so-called embedding dimension of $d = 4096$. 
Each new token introduces one additional embedding vector of dimension $d$, as well as one additional row in the tied output projection matrix (LM head). 
Hence, the parameter increase per token is 
\[
2 \times d = 2 \times 4096 = 8192.
\]
With $n = 266$ new tokens, the total number of additional parameters is 
\[
n \times 8192 = 266 \times 8192.
\]
Carrying out the multiplication,
\[
266 \times 8192 = 2{,}179{,}072 \approx 2.18 \text{M}.
\]
Thus, extending the vocabulary with 266 tokens induces approximately $2.18$ million new parameters in LLaMA-8B. Relative to the full LLaMA-8B model, which contains approximately 
$8 \times 10^{9}$ parameters, the increase is
\[
\frac{2.179 \times 10^{6}}{8 \times 10^{9}} 
\approx 2.72 \times 10^{-4} \approx 0.027\%.
\]
Hence, extending the vocabulary by 266 tokens adds only about $0.03\%$ to the total parameter count of LLaMA-8B, a negligible increase.
\\

\begin{table}[t]
\centering
\caption{Comparison between GroverGPT and GroverGPT+.}
\label{tab:grovergpt_comparison}
\begin{tabular}{@{} l l l @{}} 
\toprule
\textbf{Aspect} & 
\textbf{GroverGPT} & 
\textbf{GroverGPT+} \\
\midrule

\textbf{Task} & 
\parbox[t]{5cm}{Approximate simulation of Grover’s algorithm outcomes} & 
\parbox[t]{5cm}{Symbolic analysis of the step-by-step quantum logic of Grover’s algorithm} \\
\midrule

\textbf{Task focus} & 
\parbox[t]{5cm}{Approximation accuracy} & 
\parbox[t]{5cm}{Interpretability and logical correctness of the reasoning trace} \\
\midrule

\textbf{Model scale} & 
\parbox[t]{5cm}{8 billion parameters} & 
\parbox[t]{5cm}{8 billion parameters} \\
\midrule

\textbf{Data scale} & 
\parbox[t]{5cm}{3--20 qubits w/o analytic trace} & 
\parbox[t]{5cm}{Analytic CoT Data, \textbf{with Full-circuit Input:} 2--7 qubits, \textbf{with Oracle-only Input:} 2--13 qubits} \\
\midrule

\parbox[t]{3cm}{\textbf{Relative parameter increase induced by extending tokenizer}} & 
\parbox[t]{5cm}{N/A} & 
\parbox[t]{5cm}{0.027\%} \\
\bottomrule
\end{tabular}
\end{table}


\section{Relation between Classical Fidelity and Quantum State Fidelity}
\label{appendix:Relation between Bhattacharyya Score and Quantum State Fidelity}

For quantum states expressed as its density matrix forms $\rho$ and $\sigma$, the fidelity is
\begin{equation}
F(\rho,\sigma)\;=\;\bigl(\mathrm{tr}\sqrt{\sqrt{\rho}\,\sigma\,\sqrt{\rho}}\bigr)^2,
\end{equation}
and for pure states $\rho=\ket{\psi}\!\bra{\psi}$, $\sigma=\ket{\phi}\!\bra{\phi}$ it reduces to
$F(\rho,\sigma)=|\braket{\psi}{\phi}|^2$.
Let $\psi_i=\braket{i}{\psi}$ and $\phi_i=\braket{i}{\phi}$ be amplitudes in the computational basis, and define
$p_i=|\psi_i|^2$, $q_i=|\phi_i|^2$.
By Cauchy–Schwarz~\citep{alzer1999cauchy},
\[
|\braket{\psi}{\phi}|
= \Bigl|\sum_i \psi_i^*\phi_i\Bigr|
\;\le\; \sum_i |\psi_i||\phi_i|
\;=\; \sum_i \sqrt{p_i q_i}\,,
\]
hence
\[
F(\rho,\sigma)\;=\;|\braket{\psi}{\phi}|^2
\;\le\; \Bigl(\sum_i \sqrt{p_i q_i}\Bigr)^2
\;=\; CF(p,q).
\]
Equality holds if and only if the component-wise relative phases are aligned in this basis, \ie, there exists a global phase $\theta$ with $\phi_i = e^{i\theta}|\phi_i|$ and $\arg\psi_i$ equal modulo $\theta$ for all $i$, in which case the Classical Fidelity (CF) coincides with the quantum state fidelity. 
In general circuits with nontrivial relative phases, CF measures \emph{distributional similarity} appropriate for tasks defined on measurement outcomes, while full state fidelity $F(\rho,\sigma)$ should be used when amplitudes are accessible.\\

\section{Parameter Efficient Fine-Tuning with Low-Rank Adaptation}
\label{sec:method-peft-lora}

When adapting the LLMs for domain-specific applications, such as quantum computing, traditional full fine-tuning methods would require updating all parameters and present significant computational overhead due to the LLMs' extensive parameter counts~\citep{devlin2019bertpretrainingdeepbidirectional, liu2019robertarobustlyoptimizedbert, deepseekai2025deepseekv3technicalreport,jiang2024mixtralexperts}, \textit{e.g.} the LLaMA-$3$~\citep{grattafiori2024llama3herdmodels} model with $8$ billion parameters used in our work. 
To alleviate this, we use a parameter-efficient fine-tuning~(PEFT) technique called Low-Rank Adaptation~(LoRA) \citep{hu2021loralowrankadaptationlarge}. 

Specifically, LoRA builds on the observation that weight updates of pre-trained LLMs typically have low intrinsic dimensionality. 
Consider a pre-trained LLM weight $\mathtt{W} \in \mathbb{R}^{d\times k}$ in the LLM. 
Rather than directly modifying $\mathtt{W}$, LoRA introduces a decomposition of the weight update:
\begin{align}
    \mathtt{W}' = \mathtt{W} + \Delta \mathtt{W} = \mathtt{W} + \mathtt{B}\mathtt{A},
\end{align}
where $\mathtt{B} \in \mathbb{R}^{d\times r}$, $\mathtt{A} \in \mathbb{R}^{r\times k}$, and rank $r\ll \min(d, k)$. During fine-tuning, we freeze the original weights $W$ and only update the significantly smaller matrices $\mathtt{A}$ and $\mathtt{B}$, which reduces the trainable parameter count from $d \times k$ to $r \times (d + k)$.

For training the model, we apply LoRA only to query and value attention modules to improve efficiency further. 
We apply supervised fine-tuning to the chain-of-thought data as introduced in Section~IV B, with a dataset $\mathcal{D} = {(x_i, y_i)}_{i=1}^{N}$, where $x_i$ is the input text and $y_i$ are the expected outputs including intermediate reasoning steps. The training objective is the standard autoregressive language model loss~\citep{brown2020languagemodelsfewshotlearners, radford2019language}:
\begin{equation}
\mathcal{L}_\text{SFT} = -\sum_{i=1}^{N} \sum_{j=1}^{|y_i|} \log p_\theta(y_{i,j} | x_i, y_{i,<j}),
\end{equation}
where $\theta$ represents only the LoRA parameters, $y_{i,j}$ is the $j$-th token of the $i$-th response, and $y_{i,<j}$ denotes the preceding tokens in the sequence.

This parameter-efficient supervised fine-tuning approach enables the processing of longer QASM and quantum chain-of-thought sequences in our work while preserving the model's general capabilities. 
Notably, it achieves comparable performance to full fine-tuning with less than $1\%$ of the trainable parameters. 

\section{The Rules of Quantum-Native Tokenization}
\label{The rules of QASM Native Tokenization}

Our quantum-native tokenizer leverages specifically designed parsing rules to systematically process and tokenize QASM circuit descriptions. This tokenizer enables accurate and efficient parsing of QASM syntax, thus significantly reducing token sequence length and computational overhead. The rules, along with their detailed implementation (provided in Supplementary Note~\ref{code:tokenizer}), are elaborated below:

\textbf{Gate Definition Parsing.} Gate definitions in QASM typically have the format:
    \[
    \texttt{gate gate\_name(parameter\_list) qubit\_list \{ }
    \]
To accurately parse these definitions, we utilize regular expressions to capture three main components:
    
\begin{itemize}
    \item \textit{Gate Name}: Extracted as a single token.
    \item \textit{Parameters}: Extracted individually if present; otherwise, this component can be empty.
    \item \textit{Target Qubits}: Qubit arguments are extracted separately, supporting multiple qubit entries.
\end{itemize}

After extracting these components, the tokenizer performs normalization by removing numerical suffixes that follow specific internal naming conventions, such as \texttt{\_gate\_q\_}, \texttt{unitary\_}, or \texttt{mcx\_vchain\_}. This normalization ensures consistency and compactness, abstracting from redundant indexing information which does not affect semantics.

\textbf{Operation Command Handling.} Standard quantum operation commands in QASM typically appear as follows:
\[
\texttt{operation\_name(parameter\_list) qubit\_list;}
\]
Each operation command is parsed by first identifying the operation name, optional parameters (when applicable), and targeted qubits separately. This parsing rule uses regular expressions to isolate and tokenize these segments. Similar to gate definition parsing, any numerical suffixes associated with the internal naming conventions are removed to achieve uniformity. This approach ensures each token reflects a meaningful semantic unit rather than arbitrary indexing.

\textbf{Bracket and Structural Delimiters.} Correctly interpreting hierarchical and nested structures in QASM descriptions is crucial, particularly when handling gate definitions and quantum circuit subroutines. To explicitly address this, our parsing rules incorporate handling of structural delimiters, such as opening (\texttt{\{}) and closing braces (\texttt{\}}). Each occurrence of such delimiters is tokenized independently, enabling accurate reconstruction and analysis of nested and hierarchical structures within complex quantum circuit definitions.

\textbf{Empty Lines and Whitespace Management.} As part of our robust tokenization approach, the tokenizer explicitly strips and ignores any empty or whitespace-only lines, ensuring only meaningful QASM commands are processed and tokenized.

\textbf{Error Handling.} If the tokenizer encounters any line that does not conform to the recognized patterns (gate definition, standard operation, or structural delimiters), a clear and informative syntax error is raised. This strict rule enforcement helps ensure the integrity and correctness of the parsing and subsequent analyses.

Collectively, these parsing rules effectively reduce redundant tokenization, maintain semantic coherence, and significantly decrease sequence length. Consequently, this enhances computational efficiency during the symbolic analysis of quantum circuits.

\begin{lstlisting}[style=quantumcode, language=Python, caption={Python implementation for the Rules of Quantum-Native Tokenization}, label=code:tokenizer]
def _tokenize_line(command):
    """Tokenizes a line of quantum assembly command into structured tokens.
    
    Args:
        command: Input command string to be tokenized
        
    Returns:
        List of tokens representing the command
    """
    command = command.strip()
    if not command:
        return []

    # Handles gate definitions (e.g., "gate h q {")
    if command.startswith("gate"):
        gate_match = re.match(r"gate\s+(\w+)(?:\s*\((.*?)\))?\s+([^{]+)\s*{", command)
        if not gate_match:
            raise SyntaxError(f"Invalid gate definition: {command}")
            
        # Extract components from gate declaration
        gate_name = gate_match.group(1)
        params_part = gate_match.group(2) or ""
        qubits_part = gate_match.group(3)
        
        # Process parameters and qubits
        params = [p.strip() for p in params_part.split(",") if p.strip()]
        qubits = [q.strip() for q in qubits_part.split(",") if q.strip()]
        
        # Generate tokens and standardize names
        tokens = ["gate", gate_name] + params + qubits + ["{"]
        tokens = [re.sub(r'^(_gate_q_|unitary_|mcx_vchain_)\d+$', r'\1', t) for t in tokens]
        return tokens

    # Handles standard operations (e.g., "h q[0];")
    groups = re.match(r"^(\w+)(?:\((.*?)\))?\s+([^;]+);", command)
    if groups:
        op_name = groups.group(1)
        params = groups.group(2)
        targets = groups.group(3)
        
        # Build token sequence
        tokens = [op_name]
        if params:
            tokens += ["("] + [p.strip() for p in params.split(",")] + [")"]
        tokens += [t.strip() for t in targets.split(",")]
        tokens = [token for token in tokens if token]
        
        # Normalize special tokens
        tokens = [re.sub(r'^(_gate_q_|unitary_|mcx_vchain_)\d+$', r'\1', t) for t in tokens]
        return tokens

    # Handles closing braces
    if command == "}":
        return ["}"]

    raise SyntaxError(f"Unrecognized command: {command}")
\end{lstlisting}

\section{Additional Details on Scalability}
\label{app:simulation_baseline}

\begin{lstlisting}[style=quantumcode, language=Python, caption={Python implementation of Unitary simulation baseline}, label=code:unitary_simulation]
def simulate_qasm(qasm_str: str) -> float:
    """Perform Unitary simulation and return elapsed time."""
    st = time.time()
    try:
        processed_code = preprocess_qasm_code(qasm_str)
        qc = parse(processed_code)
        qc.remove_final_measurements(inplace=True)
        
        unitary = Operator(qc).data  
        initial_state = np.zeros(2**qc.num_qubits)
        initial_state[0] = 1
        final_state = unitary @ initial_state
        probabilities = np.abs(final_state)**2
        num_qubits = qc.num_qubits
        states = [f"{i:0{num_qubits}b}" for i in range(2**num_qubits)]
        for state, prob in zip(states, probabilities):
            print(f"|{state}>: {prob:.10f}")
    except Exception as e:
        print(f"Simulation failed: {str(e)}")
        return 0.0
    return time.time() - st
\end{lstlisting}

\begin{lstlisting}[style=quantumcode, language=Python, caption={Python implementation of SV simulation baseline}, label=code:statevector_simulation]
def simulate_qasm(qasm_str: str) -> float:
    """Perform Statevector simulation and return elapsed time."""
    st = time.time()
    try:
        processed_code = preprocess_qasm_code(qasm_str)
        qc = parse(processed_code)
        qc.remove_final_measurements(inplace=True)
        statevector = Statevector(qc)
        probabilities = np.abs(statevector.data)**2
        num_qubits = qc.num_qubits
        states = [f"{i:0{num_qubits}b}" for i in range(2**num_qubits)]
        for state, prob in zip(states, probabilities):
            print(f"|{state}>: {prob:.10f}")    
    except Exception as e:
        print(f"Simulation failed: {str(e)}")
        return 0.0
    return time.time() - st
\end{lstlisting}

\begin{lstlisting}[style=quantumcode, language=Python, caption={Python implementation of DM simulation baseline}, label=code:densitymatrix_simulation]
def simulate_qasm(qasm_str: str) -> float:
    """Perform Density Matrix simulation and return elapsed time."""
    st = time.time()
    try:
        processed_code = preprocess_qasm_code(qasm_str)
        qc = parse(processed_code)
        qc.remove_final_measurements(inplace=True)
        density_matrix = DensityMatrix(qc)
        probabilities = np.real(np.diag(density_matrix.data))
        num_qubits = qc.num_qubits
        states = [f"{i:0{num_qubits}b}" for i in range(2**num_qubits)]
        for state, prob in zip(states, probabilities):
            print(f"|{state}>: {prob:.10f}")
    except Exception as e:
        print(f"Simulation failed: {str(e)}")
        return 0.0
    return time.time() - st
\end{lstlisting}

\begin{figure}[h!]
\centering
\includegraphics[width=0.48\textwidth]{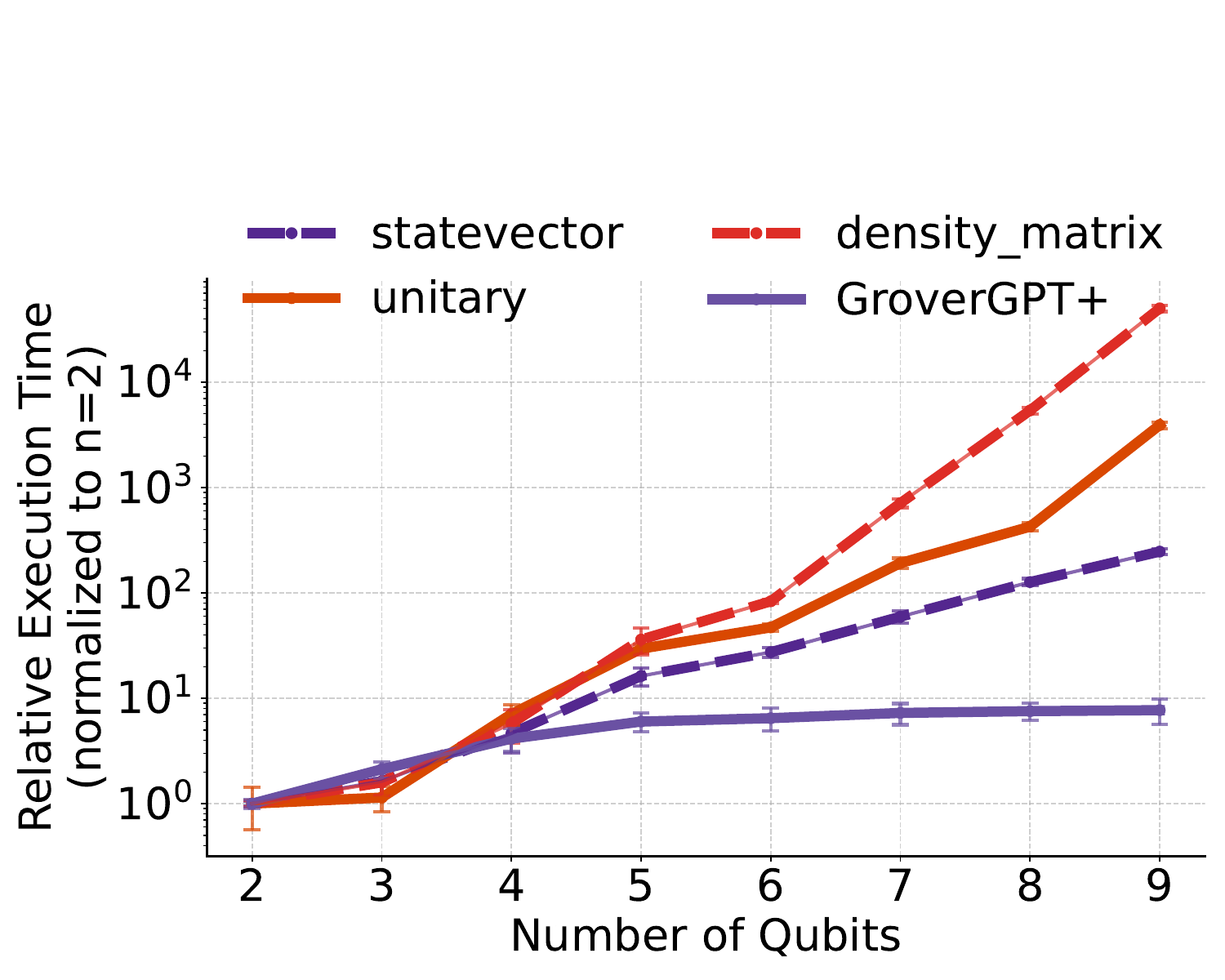}
\vspace{-0.5cm}
\caption{Relative execution time of GroverGPT+'s symbolic analysis task compared to traditional classical simulation methods—State Vector, Unitary, and Density Matrix simulations—across different numbers of qubits. All values are normalized to the execution time at 2 qubits. Solid lines show means; discrete error bars indicate uncertainty (mean $\pm$\, std).}
\label{fig:computational_advantage}
\end{figure}

In this section, we provide a detailed comparison of the \emph{computational scalability} of GroverGPT+ against traditional simulation baselines in terms of end-to-end latency, as referenced in Section~II E (Please refer to this section for the motivation of this comparison). We compare both relative latency, which highlights scaling trends, and absolute latency, which provides a concrete performance reference. Below we first introduce the baselines and task specifications.

\textbf{(i) Baselines.} We choose some traditional simulation approaches as the baselines, including State Vector (SV), Unitary, and Density Matrix (DM) simulations. Specifically, the SV simulation maintains a length-$2^n$ complex vector and updates it gate by gate, with time complexity $\mathcal{O}(G \cdot 2^n)$ for $G$ gates. This is the standard exact simulation method for pure states. The Unitary simulation explicitly constructs the full $2^n \times 2^n$ unitary matrix corresponding to the circuit and applies it to the initial state. This requires at least $\mathcal{O}(4^n)$ time for a single matrix–vector multiplication, and is therefore only practical for small $n$, serving as a reference implementation. The DM simulation computes the final density matrix $\rho \in \mathbb{C}^{2^n \times 2^n}$ from which the output probabilities are extracted from the diagonal entries. This method naturally generalizes to noisy or mixed-state settings but incurs even higher cost as $\mathcal{O}(G \cdot 8^n)$ time. Detailed code implementations are provided, see Supplementary Note~\ref{code:unitary_simulation}, Supplementary Note~\ref{code:statevector_simulation} and Supplementary Note~\ref{code:densitymatrix_simulation}.

\textbf{(ii) Task Specifications.} For this comparison, it is crucial to recognize that the methods perform fundamentally different tasks. The classical baselines execute \textbf{numerical state evolution}, a purely computational process to calculate the final state vector. In contrast, GroverGPT+ performs \textbf{symbolic analysis}, generating a reasoning trace that interprets the circuit's logic and infers the final probabilities from that understanding. Although the tasks differ, they share a common input format (QASM) and a comparable final output (a probability distribution), which makes a comparison of their end-to-end latency a useful, albeit indirect, point of reference.

\textbf{(iii) Configurations.} The experiments are conducted by measuring the execution time of each method across qubit sizes ranging from 2 to 9. For each method, three runs are performed to compute the mean and standard deviation of execution time. For GroverGPT+, the reported runtimes refer to inference time only and do not include training.

\textbf{Relative Computational Scalability.} Results are similarly plotted on a logarithmic scale to highlight scalability differences. As shown in Supplementary Figure~\ref{fig:computational_advantage}, classical methods exhibit steep growth in execution time with increasing qubit numbers. For example, DM and Unitary simulations show sharp exponential scaling, with relative execution times surpassing \(10^2\) by 7 qubits. In contrast, GroverGPT+ displays a notably gentler slope, maintaining sub-linear growth. Instead of performing tensor-based state evolution, GroverGPT+ generates output probability distributions from its symbolic analysis of the QASM input via CoT reasoning, thereby avoiding the exponential computational overhead inherent to matrix-based simulations. While GroverGPT+ does not defeat exponential complexity in the worst case, it effectively trades offline training for more favorable runtime scaling on its targeted analysis task.

\begin{remark}
    \textit{Due to the inherent overhead of large language models, the absolute runtime of GroverGPT+ on small qubit instances is indeed larger than that of classical simulators. Therefore, we do not claim an absolute runtime advantage. Instead, we focus on the empirical \emph{scaling trend} of GroverGPT+, which, in the case of Grover’s algorithm, exhibits a surprisingly gentle growth compared to the steep exponential increase of classical methods. This is intriguing as it suggests a complementary perspective on runtime scaling beyond traditional simulation techniques.}
\end{remark}

\begin{figure}[t!]
\centering
\includegraphics[width=0.4\textwidth]{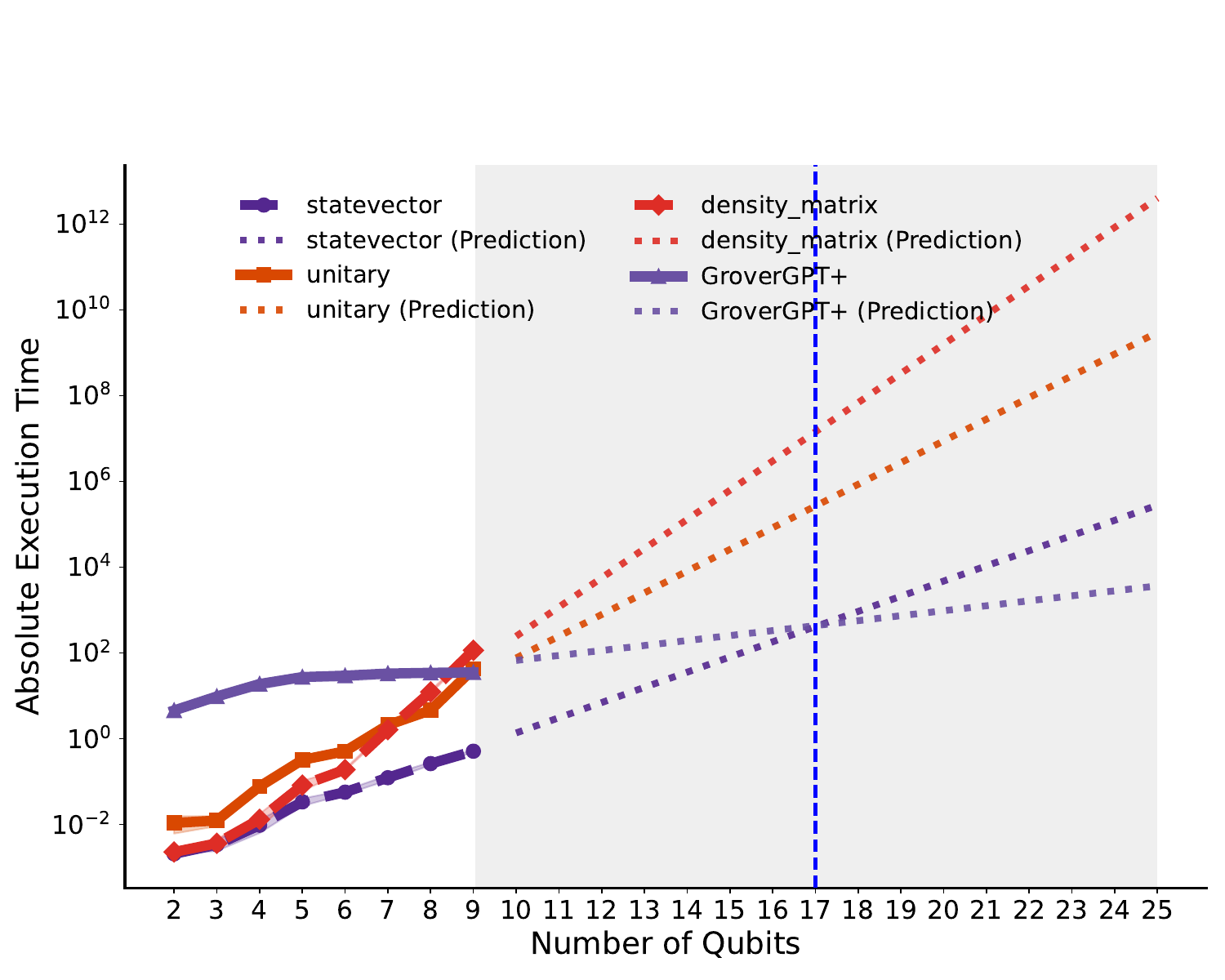}
\caption{
    Absolute execution time of GroverGPT+ and traditional classical simulation methods, with an extrapolation to visualize the implications of their different scaling trends. Solid lines correspond to measured data (2--9 qubits); dashed lines show extrapolated predictions based on measured data fitting. In the legend, the term \textbf{``Prediction''} represents the extrapolated portions of the curves.}
\label{fig:extrapolate}
\end{figure}

\textbf{Absolute Computational Scalability.} Beyond the measured data, we compare the absolute execution time and perform an extrapolation to visualize where the different scaling trends might lead, see Supplementary Figure~\ref{fig:extrapolate}. In this figure, solid lines show the scaling of measured data points (2--9 qubits), while dashed lines represent extrapolations (predictions) based on the observed data fitting. For clarity of visualization, we do not include shaded regions indicating standard deviation for the curves of measured data, as it does not influence the scaling trend. See the standard deviation of the curves of measured data in Supplementary Figure~\ref{fig:computational_advantage}. As for the extrapolated portions, the predicted curves are not actual measurements and do not have associated uncertainty estimates from empirical data. Based on these trends observed within the tested range (2--9 qubits), a \textbf{crossover point}, where the latency of GroverGPT+ might hypothetically become favorable compared to the SV simulator, is projected to occur around 17 qubits. We emphasize that this projection is purely speculative and should not be interpreted as a performance claim or guarantee. We present this extrapolation not as a definitive prediction, but as an illustration of the long-term implications of the different scaling behaviors observed within our tested range (2--9 qubits). Notably, this extrapolation for GroverGPT+ is conservative, as it assumes log-linear growth, whereas the observed data in Figure 6 in the main text and Supplementary Figure~\ref{fig:computational_advantage} suggests a sub-linear trend within the tested regime. However, we cannot guarantee that this trend will continue beyond the tested range, and the crossover projection should be viewed as a hypothetical scenario rather than an empirical finding.

Below we detail the hardware and inference setup, relation related to training size and storage and token-length to time complexity and context limitations regarding these comparisons.

\noindent \textbf{i) Hardware and Inference Setup.}
All classical simulator baselines are executed on a MacBook Pro equipped with the Apple M4 Pro SoC (14-core CPU, 20-core GPU, and 24 GB unified memory). GroverGPT+ inference is performed on an NVIDIA RTX A6000 GPU with 48 GB GDDR6 VRAM. We do not apply batching or parallel inference; all runtimes are measured in single-input, single-instance inference mode.

\noindent \textbf{ii) On Training-set Size and Storage.}
Reported runtimes refer strictly to inference. We do not include training time or training dataset storage in the benchmarks. While the training corpus has a non-trivial size, its storage footprint does not affect inference latency.

\noindent \textbf{iii) Token-length to Time Complexity and Context Limitations.} GroverGPT+ inference time scales approximately linearly with the number of input tokens. Full-circuit QASM inputs grow rapidly in token length (approximately exponential in qubit count), which encounters the model’s context-length bottleneck. By contrast, the Oracle-only input format reduces token growth to a more manageable rate, leading to a milder increase in runtime.

\section{Complete Analysis Process via Chain-of-Thought Reasoning}
\label{Complete Chain-of-Thought Reasoning Process}

\begin{tcolorbox}[colback=blue!5!white, colframe=gray!75!black, title=Supplementary Note 7: Chain-of-Thought Analysis Output Analysis, label=code:complete_cot]

\textbf{=== Analysis ===}

The Oracle entity is extracted below:
\begin{verbatim}
  x _gate_q_3;
  mcmt _gate_q_0, _gate_q_1, _gate_q_2, _gate_q_3;
  x _gate_q_3;
  x _gate_q_1;
  mcmt _gate_q_0, _gate_q_1, _gate_q_2, _gate_q_3;
  x _gate_q_1;
\end{verbatim}

\textbf{=== Block 1 ===} \\
Operation sequence:
\begin{verbatim}
x _gate_q_3;
mcmt _gate_q_0, _gate_q_1, _gate_q_2, _gate_q_3;
x _gate_q_3;
\end{verbatim}
State construction: \\

x \_gate\_q\_0: Absent → 1, then → 1 \\
x \_gate\_q\_1: Absent → 1, then → 11 \\
x \_gate\_q\_2: Absent → 1, then → 111 \\
x \_gate\_q\_3: Present → 0, then → 0111 \\

Final state: \texttt{0111}

\textbf{=== Block 2 ===} \\
Operation sequence:
\begin{verbatim}
x _gate_q_1;
mcmt _gate_q_0, _gate_q_1, _gate_q_2, _gate_q_3;
x _gate_q_1;
\end{verbatim}
State construction:\\

x \_gate\_q\_0: Absent → 1, then → 1\\
x \_gate\_q\_1: Present → 0, then → 01\\
x \_gate\_q\_2: Absent → 1, then → 101\\
x \_gate\_q\_3: Absent → 1, then → 1101\\

Final state: \texttt{1101}

\textbf{=== Final Marked States ===} \\
\texttt{0111}\\
\texttt{1101} 

\textbf{=== Results ===}
\begin{verbatim}
{
 '0111': 0.4727,
 '1101': 0.4727,
 '0000': 0.0039,
 '0001': 0.0039,
 '0010': 0.0039,
 '0011': 0.0039,
 '0100': 0.0039,
 '0101': 0.0039,
 '0110': 0.0039,
...
}
\end{verbatim}

\end{tcolorbox}

In this section, we present the complete analysis workflow for Grover's algorithm based on GroverGPT+'s CoT reasoning paradigm. Supplementary Note 7 provides an example of the CoT process. Through this example, we draw some conclusions regarding the capability of LLMs as tools for the symbolic analysis of quantum algorithms:

\begin{conclusion}
    An LLM can effectively perform entity extraction from the quantum circuit representation of an algorithm.
\label{conclude:entity_extract}
\end{conclusion}

This conclusion highlights a non-trivial capability of LLMs beyond conventional natural language tasks. In the context of quantum circuit analysis, entity extraction refers to identifying structured and meaningful components (e.g., oracle constructions) from a sequence of low-level QASM. GroverGPT+ demonstrates the ability to parse quantum programs, including those with nested gate definitions, parameterized operations, and qubit registers. Beyond syntactic parsing, it is capable of understanding the semantic structure of circuits, recognizing functionally coherent blocks, such as the Oracle, that are essential to the algorithm's logic. Furthermore, GroverGPT+ can localize and extract specific substructures (entities) from quantum circuits purely based on its learned CoT reasoning, even in the absence of explicit markers or prompts. This suggests that LLMs trained with appropriately designed CoT steps can generalize entity extraction to structured QASM-like inputs.

\begin{conclusion} 
An LLM can effectively distinguish qubit operations associated with marking different target states based on the sequential structure of qubit manipulations. 
\label{conclude:marked_state}
\end{conclusion}

In our application, this capability is critical for accurately identifying the marked states, which in turn enables GroverGPT+ to output the corresponding probability amplitudes with high fidelity. As illustrated in Supplementary Note 7, this conclusion captures how GroverGPT+ successfully locates the marked states, as detailed below:

First, GroverGPT+ is able to decompose the full oracle construction into sub-regions, referred to as \textit{blocks}, where each block corresponds to the operations defining a specific marked state. Second, within each block, GroverGPT+ identifies the sequence of single-qubit operations leading up to the application of the multi-controlled multi-target (MCMT) gate. By systematically analyzing the presence (0) or absence (1) of $X$ gates on each qubit, GroverGPT+ incrementally constructs the corresponding computational basis state. The final constructed string (e.g., \texttt{0111} as shown in Supplementary Note~7) is recognized as the marked state associated with the block's operations. This demonstrates the model’s ability to not only trace quantum operations but also to semantically translate operational sequences into measurable quantum states.

Moreover, in the Qiskit implementation of oracle constructions, bit-string reversal typically occurs twice: first, explicitly within the oracle to match Qiskit's qubit indexing convention; and second, implicitly during final state measurement to restore the original bit-order. To streamline the analysis process, GroverGPT+ effectively consolidates these two reversal steps into a single operation, thereby eliminating redundancy and significantly enhancing reasoning efficiency.

\begin{conclusion}
    An LLM can learn an effective mapping from its symbolic analysis to the algorithm's correct output probability distribution.
\label{conclude:output}
\end{conclusion}


After completing the symbolic analysis to identify key parameters like the number of qubits ($n$) and the number of marked states ($t$), GroverGPT+ must infer the final probabilities. This is not a numerical simulation of state evolution, but rather an approximation of a complex mapping that implicitly accounts for factors such as the optimal number of Grover iterations. Analytically, the outcome probabilities for both marked and unmarked states in Grover's algorithm can be determined based on the following mathematical formulation:

Given an \( n \)-qubit quantum system, the total number of computational basis states is \( N = 2^n \). Suppose there are \( t \) marked states within the system. The initial amplitude angle \( \theta \) is defined as:
\begin{equation}
    \theta = \arcsin\left(\sqrt{\frac{t}{N}}\right).
\end{equation}
The optimal number of Grover iterations \( k_{\text{opt}} \) that maximizes the success probability is approximately:
\begin{equation}
    k_{\text{opt}} = \left\lfloor \frac{\pi}{4}\sqrt{\frac{N}{t}} \right\rfloor.
\end{equation}
After performing \( k_{\text{opt}} \) iterations, the probability \( P_{\text{marked}} \) of measuring one of the marked states is given by:
\begin{equation}
    P_{\text{marked}} = \sin^2\left((2k_{\text{opt}} + 1)\theta\right),
\end{equation}
while the probability \( P_{\text{unmarked}} \) of measuring any unmarked state is:
\begin{equation}
    P_{\text{unmarked}} = \cos^2\left((2k_{\text{opt}} + 1)\theta\right).
\end{equation}
Thus, based on the number of qubits and the number of marked states identified during the analysis, one can analytically calculate the expected output distribution of Grover's algorithm.

Leveraging its high representational capacity, GroverGPT+ is capable of approximating this complex input–output mapping. Based on the number of qubits, the number of extracted blocks (each corresponding to a marked state), and the identified marked states, GroverGPT+ assigns output probabilities for both marked and unmarked states without explicitly performing quantum evolution. This mapping is learned via parameter-efficient supervised fine-tuning (see \textit{Supplementary Note~\ref{sec:method-peft-lora}} for more details).

Notably, while GroverGPT+ can generalize this mapping, a slight fidelity loss may occur when analyzing unseen configurations not present in the training data. Empirical results show that GroverGPT+ successfully captures the key characteristics of probability amplitude distributions, distinguishing marked states from unmarked states. Nevertheless, without explicitly calculating the underlying quantum amplitudes, the predicted outputs may exhibit minor deviations compared to ground truth results from classical simulators, particularly when extrapolating to unseen scenarios. 

\begin{conclusion}
    Tokenization requires task-specific and cross-domain-aware designs to achieve scalability, memory efficiency, and compatibility with mixed natural and quantum language outputs.
\end{conclusion}

\begin{figure*}
    \centering
    \includegraphics[width=\textwidth]{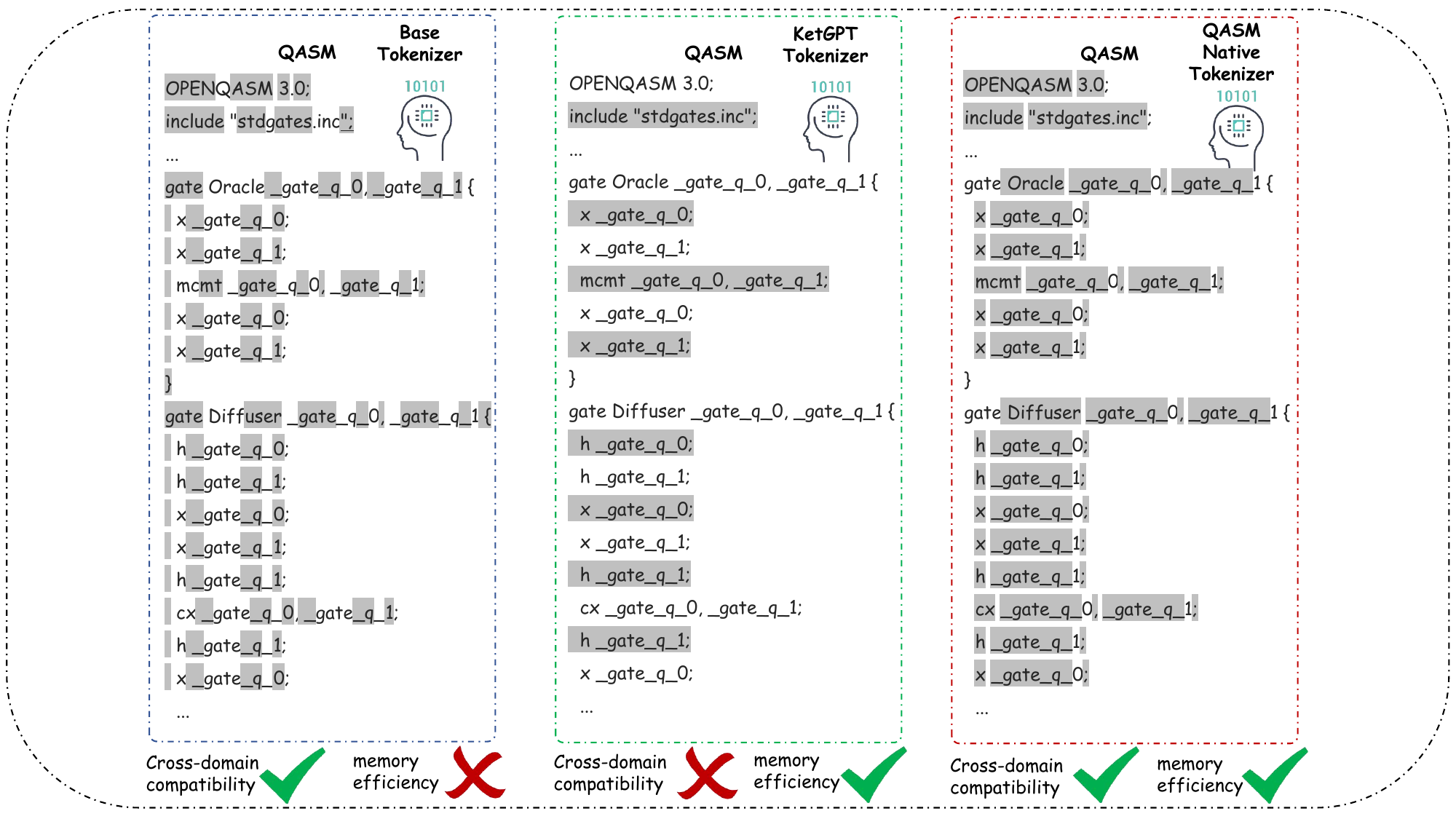}    
    \vspace{-10pt}
    \caption{Comparison of tokenization strategies on a sample QASM snippet across three tokenizers: LLaMA-3 base tokenizer, KetGPT tokenizer, and the proposed quantum-native tokenizer. Each grey and non-grey segment represents a distinct token. The LLaMA-3 tokenizer demonstrates cross-domain compatibility, applicable to both natural language and QASM domains; however, its non-domain-specific token splitting leads to fragmented semantics and significantly increases input context length, thus consuming more GPU memory. The KetGPT tokenizer effectively reduces the token sequence length and optimizes GPU memory by tokenizing each QASM line as a single token, but it lacks compatibility with natural language and suffers from rapidly increasing model parameters and potential under-training as the QASM length grows. The quantum-native tokenizer, our proposed method, is designed to address these limitations effectively by balancing token granularity, memory efficiency, and domain compatibility.}
    \label{fig:tokenizer_compare}
\end{figure*}

The design of a tokenizer significantly impacts the efficiency and effectiveness of downstream tasks, especially when dealing with domain-specific representations, such as QASM. As demonstrated in Supplementary Figure~\ref{fig:tokenizer_compare}, three tokenizers, including the base tokenizer, KetGPT~\citep{apak2024ketgpt}'s tokenizer and our quantum-native tokenizer, are compared. KetGPT's tokenizer treats each line of QASM as an individual token ID, facilitating efficiency in the data augmentation task. However, KetGPT's tokenizer may exhibit several limitations when applied to our task as detailed below:

Firstly, KetGPT's line-level tokenization scheme specifically tailors for tasks where both inputs and outputs are purely QASM-based, overlooking the challenges posed by cross-domain tasks where natural language and quantum programming syntax are interleaved. Using a coarse-grained tokenizer for QASM risks interfering with natural language generation. A finer-grained tokenization for QASM elements is therefore essential for enabling flexible decoding across heterogeneous domains. Secondly, it tokenizes entire QASM lines as atomic units. While this may suffice for tasks purely centered on quantum circuit augmentation, it leads to potentially infinite combinations of gate types, qubit indices, and parameters. Consequently, the tokenizer's vocabulary would expand uncontrollably, significantly increasing both the size of the model’s embedding table and the total number of trainable parameters. This expansion may result in under-training of the model, as the required data size must scale accordingly with the growth in trainable parameters~\citep{vapnik2015uniform}, limiting the model’s scalability in practice. To better demonstrate this point, we aim to measure the increase in trainable parameters. Since the size of the embedding layer is directly proportional to the vocabulary size, it suffices to compare the increase in vocabulary size. The statistical analysis is shown below:

We generate a corpus of QASM circuit descriptions with varying numbers of qubits, ranging from 2 to 9 qubits. For each qubit number, we apply both tokenization strategies to the corpus and measure the resulting vocabulary size growth after tokenizing all circuits. The results are presented in Supplementary Figure~\ref{fig:vocab_growth}. 

\begin{figure*}[!h]
\centering
\includegraphics[width=0.6\textwidth]{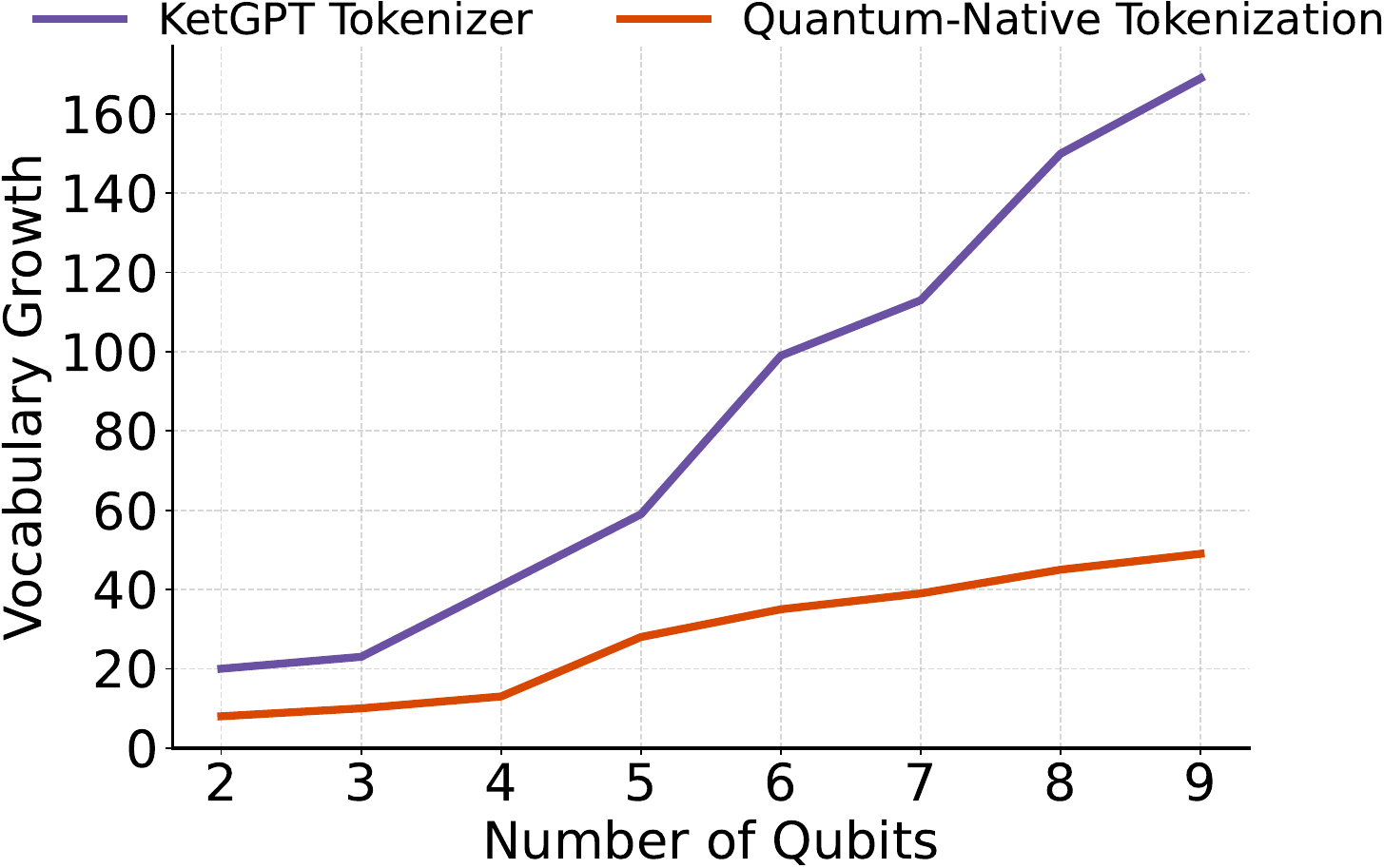}
\vspace{-0.3cm}
\caption{Vocabulary growth as a function of the number of qubits under different tokenization strategies.}
\label{fig:vocab_growth}
\end{figure*}

Accordingly, there are several key observations:  

The vocabulary size under our quantum-native tokenizer grows slowly with the number of qubits. This reflects the effectiveness of the fine-grained design in capturing the compositional structure of QASM without inducing vocabulary explosion. In contrast, the KetGPT tokenizer shows faster vocabulary growth, driven by the combinatorial increase in unique line patterns as circuit complexity rises. Hence, treating entire QASM lines as atomic tokens may inflate the model’s embedding size significantly.

Therefore, combining the analysis for the comparison between the base tokenizer and quantum-native tokenization in Section IV A, designing a domain-aware tokenizer that captures the semantic structures of QASM while being compatible with natural language modeling is crucial. This conclusion highlights that tokenization is not a mere preprocessing step but a foundational design choice.

\section{CoT Training Details}
\label{appendix:CoT_training_details}

CoT reasoning is an emergent capability in LLMs that enables them to reason through complex problem-solving tasks by generating step-by-step reasoning paths~\citep{wei2022emergentabilitieslargelanguage,wei2023chainofthoughtpromptingelicitsreasoning, deepseekai2025deepseekr1incentivizingreasoningcapability}. 
Techniques like appending ``\textit{Let's think step by step}" to prompts \citep{kojima2023largelanguagemodelszeroshot} or providing in-context examples \citep{nachane2024shotchainofthoughtdrivenreasoning} can elicit CoT reasoning. 
Advanced methods such as Self-Consistency \citep{wang2023selfconsistencyimproveschainthought}, Tree-of-Thought \citep{yao2023treethoughtsdeliberateproblem}, and Least-to-Most prompting \citep{zhou2023leasttomostpromptingenablescomplex} further improve CoT performance. 

\textbf{CoT Training for GroverGPT+.} \citet{yao2025unveiling} analyzes the advantages and underlying mechanisms of explicit CoT training~\citep{wei2022chain}, which refers to the training strategy of explicitly incorporating intermediate reasoning steps before producing the final output. This approach has been shown to significantly improve the reasoning ability and generalization performance of LLMs, especially in tasks that require multi-step logical inference. 

In the context of quantum algorithm analysis, CoT training similarly plays a critical role. Unlike the previous approach~\citep{wang2024grovergpt} that directly predicts the final output probabilities from the quantum circuit, we explicitly model the intermediate analysis process through structured reasoning chains. These chains guide the model to generate step-by-step approximations or logical justifications for the predicted outcome. Our implementation of CoT training is primarily based on supervised fine-tuning~\citep{yue2023mammoth, yu2023metamath}, where high-quality CoT annotations are generated using classical simulators and then used to fine-tune the model (See Supplementary Figure~\ref{fig:tokenizer_compare}). This allows the LLM to internalize the procedural knowledge of quantum circuit analysis. 

Furthermore, in \textit{Supplementary Note~\ref{Full Ablation Study}}, we conduct a detailed ablation study to evaluate the effectiveness of each CoT component. The results validate that reasoning chains contribute significantly to the model's accuracy across different circuit depths and oracle configurations, which aligns with the conclusions emphasized in \citet{yao2025unveiling}: CoT training emerges as a key enabler for the generalization capabilities observed in the GroverGPT+, particularly in its ability to extrapolate to unseen configurations and maintain consistency in output distributions. In our setting, where the input is pure QASM and no additional natural language prompt is provided, CoT training acts as the backbone for aligning symbolic input understanding with semantic reasoning trajectories.

When applying the CoT training technique, we design two types of CoT data:

\begin{itemize}
    \item \textbf{CoT Data with Oracle-only Input}: This dataset contains inputs consisting solely of oracle definitions, with CoT reasoning processes provided as outputs. It is specifically designed to enable GroverGPT+ to acquire the capability of inferring marked states directly from oracle descriptions. This type of training dataset utilized in Section II D in the main text spans qubit numbers \(n \in \{2, 3, \dots, 10\}\), ensuring sufficient coverage across small- and medium-scale circuits.

    \item \textbf{CoT Data with Full-circuit Input}: This dataset includes complete QASM circuit descriptions of Grover's algorithm as inputs, paired with CoT reasoning processes as outputs. Its purpose is to guide GroverGPT+ to concentrate specifically on oracle construction and simultaneously enhance its capability. The qubit range for this dataset is restricted to \(n \in \{2, 3, \dots, 7\}\), due to the increased token length and complexity of full-circuit inputs.
\end{itemize}

The data configurations of the qubit range are consistent in all experiments reported in Section II D in the main text width, ensuring consistency across different evaluation settings. In \textit{Supplementary Note \ref{Full Ablation Study}}, we show that both types of data are indispensable in our training strategy.

\section{Full Ablation Study}
\label{Full Ablation Study}

This section conducts a full ablation study to systematically examine the contribution of each CoT component and input strategy in the symbolical analysis of Grover's algorithm with GroverGPT+. We specifically verify two major CoT reasoning modules, each corresponding to a key conclusion discussed in \textit{Supplementary Note~\ref{Complete Chain-of-Thought Reasoning Process}}:

\begin{itemize}
    \item \textbf{Entity Extraction Module} (corresponding to \textbf{Conclusion~\ref{conclude:entity_extract}}): extracting the Oracle entity from the QASM circuit.
    \item \textbf{State Construction Module} (corresponding to \textbf{Conclusion ~\ref{conclude:marked_state}}): constructing the marked computational basis states from sequential qubit operations.
\end{itemize}

In addition, we evaluate the two input types introduced in Section IV B in the main text 
\begin{itemize}
    \item \textbf{CoT Data with Oracle-only Input}: using only the Oracle subcircuit as input.
    \item \textbf{CoT Data with Full-circuit Input}: using the full Grover's algorithm circuit as input.
\end{itemize}

In each ablation experiment, we remove one CoT module or input type from the training process, keeping all other settings identical. We then evaluate the resulting model behavior by analyzing its generated outputs under the standard label format described below.

\begin{tcolorbox}[colback=blue!5!white, colframe=gray!75!black, title=Supplementary Note 8: Ablation Label: Without Entity Extraction, label=Without Entity Extraction]
\textbf{=== Block 1 ===} \\
Operation sequence:
\begin{verbatim}
x _gate_q_1;
mcmt _gate_q_0, _gate_q_1, _gate_q_2, _gate_q_3;
x _gate_q_1;
\end{verbatim}
State construction:\\

x \_gate\_q\_0: Absent → 1, then → 1\\
x \_gate\_q\_1: Present → 0, then → 01\\
x \_gate\_q\_2: Absent → 1, then → 101\\
x \_gate\_q\_3: Absent → 1, then → 1101\\

Final state: \texttt{1101}

\textbf{=== Final Marked States ===} \\
\texttt{0111}\\
\texttt{1101} 

\textbf{=== Results ===}
\begin{verbatim}
{
 '0111': 0.4727,
 '1101': 0.4727,
...
}
\end{verbatim}
\end{tcolorbox}

\begin{tcolorbox}[colback=blue!5!white, colframe=gray!75!black, title=Supplementary Note 9: Without State Construction,label=Without State Construction,]
\textbf{=== Analysis ===}

The Oracle entity is extracted below:
\begin{verbatim}
x _gate_q_3;
mcmt _gate_q_0, _gate_q_1, _gate_q_2, _gate_q_3;
x _gate_q_3;
...
\end{verbatim}

\textbf{=== Results ===}
\begin{verbatim}
{
 '0111': 0.4727,
 '1101': 0.4727,
 '0000': 0.0039,
 '0001': 0.0039,
 '0010': 0.0039,
 '0011': 0.0039,
 '0100': 0.0039,
 '0101': 0.0039,
...
}
\end{verbatim}
\end{tcolorbox}

Following the above setup, we show a representative output generated by GroverGPT+ under each ablation condition as below:

\begin{tcolorbox}[
    colback=yellow!10,     
    colframe=black,        
    coltitle=white,        
    colbacktitle=black,    
    title=Supplementary Note 10: An example of the outputs when training without the Entity Extraction Module, 
    fonttitle=\bfseries,   
    sharp corners,         
    boxrule=0.8pt,
    label=abla_Entity_Extraction_Module
]

=== Analysis ===



=== Block 1 ===  \\                                                                                   
Operation sequence: \textcolor{red}{(Incorrect)} \\                                                                                 
x \_gate\_q\_1; \\                                                                              
x \_gate\_q\_2; \\                                                                                     
x \_gate\_q\_3; \\
x \_gate\_q\_4; \\
x \_gate\_q\_5; \\ 
x \_gate\_q\_6; \\
x \_gate\_q\_7; \\
h \_gate\_q\_0;  \textcolor{red}{(H gate should not exist in the Oracle definition)} \\
x \_gate\_q\_1; \\                                                                              
x \_gate\_q\_2; \\                                                                                     
x \_gate\_q\_3; \\
...

\end{tcolorbox}

\begin{tcolorbox}[
    colback=yellow!10,     
    colframe=black,        
    coltitle=white,        
    colbacktitle=black,    
    title=Supplementary Note 11: An example of the outputs when training without the State Construction Module, 
    fonttitle=\bfseries,   
    sharp corners,         
    boxrule=0.8pt,
    label=abla_state_construction
]

=== Analysis === \\

The Oracle entity is extracted below: \textcolor{green}{(Correct)} \\

  x \_gate\_q\_0;\\
  mcmt \_gate\_q\_0, \_gate\_q\_1, \_gate\_q\_2, \_gate\_q\_3;\\
  x \_gate\_q\_0;\\

=== Final Marked States ===\\
1100 \textcolor{red}{(Though the Oracle is extracted correctly, the searched marked state is  incorrect due to the missing of the state construction module)}\\

===  Results ===\\
\textcolor{red}{(Low Fidelity)}
\begin{verbatim}
{
'1100': 0.9613,
 '0000': 0.0026,
 '0001': 0.0026,
 '0010': 0.0026,
...
}
\end{verbatim}

\end{tcolorbox}

\begin{tcolorbox}[
    colback=yellow!10,     
    colframe=black,        
    coltitle=white,        
    colbacktitle=black,    
    title=Supplementary Note 12: An example of the outputs when training without the CoT Data with Full-circuit Input, 
    fonttitle=\bfseries,   
    sharp corners,         
    boxrule=0.8pt,
    label=abla_full_circuit_input
]

\textcolor{red}{(Incorrect due to not being capable of identifying the \textcolor{red}{analysis} task)} \\

The infamous "spaghetti code"! \\

This is a very long and complex sequence of gates (logic operations) in the Q\# programming language, which is used for quantum computing. I'll try to break it down and provide some insights: \\

**Gates** \\

The sequence is composed of various gates, which are the basic building blocks of quantum algorithms. Some common gates include: \\

* `H`: Hadamard gate, which applies a specific transformation to a qubit (quantum bit). \\
* `CX`: Controlled-NOT gate, which applies a NOT operation to a target qubit if the control qubit is in
the state 1. \\
* `MEASURE`: Measures the state of a qubit, collapsing the superposition to either 0 or 1. \\
* `RESET`: Resets the state of a qubit to 0. \\

...
\end{tcolorbox}

\begin{tcolorbox}[
    colback=yellow!10,     
    colframe=black,        
    coltitle=white,        
    colbacktitle=black,    
    title=Supplementary Note 13: An example of the outputs when training without the CoT Data with Oracle-only Input, 
    fonttitle=\bfseries,   
    sharp corners,         
    boxrule=0.8pt,
    label=abla_oracle_input_input
]

=== Analysis ===

The Oracle entity is extracted below:  \textcolor{green}{(Correct)} \\

x \_gate\_q\_1;\\                                           
x \_gate\_q\_2;\\                                                                                    
x \_gate\_q\_7;\\                                                                                    
x \_gate\_q\_0, \_gate\_q\_1, \_gate\_q\_2, \_gate\_q\_3, \_gate\_q\_4, \_gate\_q\_5, \_gate\_q\_6, \_gate\_q\_7; \\         
x \_gate\_q\_1; \\                                                                                      
x \_gate\_q\_2; \\                                                                                   
x \_gate\_q\_7; \\                                                                                     
... \\

=== Block 1 ===  \\                                                                                   
Operation sequence: \textcolor{red}{(Incorrect)} \\                                                                                 
x \_gate\_q\_1; \\                                                                              
x \_gate\_q\_2; \\                                                                                     
x \_gate\_q\_7; 

\textcolor{orange}{x \_gate\_q\_0, \_gate\_q\_1, \_gate\_q\_2, \_gate\_q\_3, \_gate\_q\_4, \_gate\_q\_5, \_gate\_q\_6, \_gate\_q\_7;} \textcolor{red}{(Missing)}

\textcolor{orange}{x \_gate\_q\_1;} \textcolor{red}{(Missing)} \\                                                                                      
\textcolor{orange}{x \_gate\_q\_2;} \textcolor{red}{(Missing)} \\                                                                                   
\textcolor{orange}{x \_gate\_q\_7;} \textcolor{red}{(Missing)} \\ 

State construction: \\

...

\end{tcolorbox}

\section{Prompt Design for Benchmarking LLMs}
\label{app:prompt}

To benchmark various LLMs, we designed the following unified prompt:

\begin{tcolorbox}[colback=blue!5!white, colframe=gray!75!black, title=Supplementary Note 14: Prompt Design for Benchmarking LLMs, label=prompt_design]
You are given a quantum circuit written in OpenQASM 3.0 that implements Grover's algorithm. Your task is to classically simulate this circuit and return
the final output probability distribution over all computational basis states after all gates and before measurement. The output should be a dictionary
where each key is a bitstring and each value is the corresponding probability, rounded to four decimal places. Sort the entries in descending order by
probability. If the total number of basis states exceeds 30, only return the top 30 most probable states. Include entries with probability 0.0000 if they
appear in the top 30.

Output format (example):

\begin{verbatim}
{
  '1010': 0.9732,
  '0011': 0.0087,
  '0110': 0.0087,
  ...
}
\end{verbatim}

Here is the QASM code for simulation:

\begin{verbatim}
OPENQASM 3.0;
include "stdgates.inc";

gate mcmt _gate_q_0, _gate_q_1 {
  cz _gate_q_0, _gate_q_1;
}
gate Oracle _gate_q_0, _gate_q_1 {
  x _gate_q_0;
  x _gate_q_1;
  mcmt _gate_q_0, _gate_q_1;
  x _gate_q_0;
  x _gate_q_1;
}
gate Diffuser _gate_q_0, _gate_q_1 {
  h _gate_q_0;
  h _gate_q_1;
  x _gate_q_0;
  x _gate_q_1;
  h _gate_q_1;
  cx _gate_q_0, _gate_q_1;
  h _gate_q_1;
  x _gate_q_0;
  x _gate_q_1;
  h _gate_q_0;
  h _gate_q_1;
}
bit[2] c;
qubit[2] q;
h q[0];
h q[1];
Oracle q[0], q[1];
Diffuser q[0], q[1];
c[0] = measure q[0];
c[1] = measure q[1];
\end{verbatim}
\end{tcolorbox}

\section{Details of General Experimental Settings}
\label{appendix:exp_setting}

We describe the general experimental settings used throughout our study. Specifically, we consider different input types, qubit numbers, and marked states to comprehensively analyze the model performance. We evaluate two types of inputs in our experiments:

- \textbf{Full-circuit Input:} The complete QASM code of Grover’s quantum search algorithm, including all gate definitions and execution sequences.

- \textbf{Oracle-only Input:} A partial QASM sequence that contains only the Oracle construction, excluding other gate definitions and the ordering of gate execution.

The design of the \textbf{Oracle-only input} serves two main purposes:  
1) This input format is used during training and enables the model to efficiently learn the reasoning steps in the CoT process for analyzing quantum circuits.  
2) Considering that LLMs are constrained by a maximum context length (measured in token IDs), the full-circuit input can easily exceed this limit, especially for circuits with a large number of qubits. In contrast, the Oracle-only input allows us to extend the qubit size while remaining within the context boundary, thus facilitating the exploration of the model's extrapolation performance (See \textit{Supplementary Note~\ref{appendix:Scaling_law}}).

Notably, for both inputs we adopt the analytic solution of the optimal iteration number $k_{\text{opt}}$, which guarantees the maximum success probability for a given $(n, t)$ configuration. 
This ensures that the output distributions are consistently defined even when the iteration number $k$ is not explicitly present in the oracle-only input for our method.

For the \textbf{Full-circuit Input}, we vary the number of marked states in \(\{1, 2, 3\}\). The settings for the number of qubits depend on specific experiments. For each configuration, we evaluate on \(\max(100, 2^n)\) randomly sampled QASM circuits. The prompting strategy used to guide the baseline LLMs is detailed in \textit{Supplementary Note~\ref{app:prompt}}.

For the \textbf{Oracle-only Input}, we use the same set of marked states, and set the number of qubits to \(n \in \{2, 3, \dots, 13\}\), allowing us to investigate the model's performance in larger-scale settings. \(\max(100, 2^n)\) evaluation samples are also used for each \(n\).

All the experiments adopt a consistent data configuration to train the GroverGPT+ model, as detailed in Section IV B in the main text. Supplementary Note~\ref{appendix:Scaling_law} provides a deeper investigation by varying the data configuration.

\section{Empirical Study of the Chain-of-Thought Length}
\label{appendix:Empirical Study of the Chain-of-Thought Length}

\begin{figure}[h!]
\centering
\includegraphics[width=0.7\textwidth]{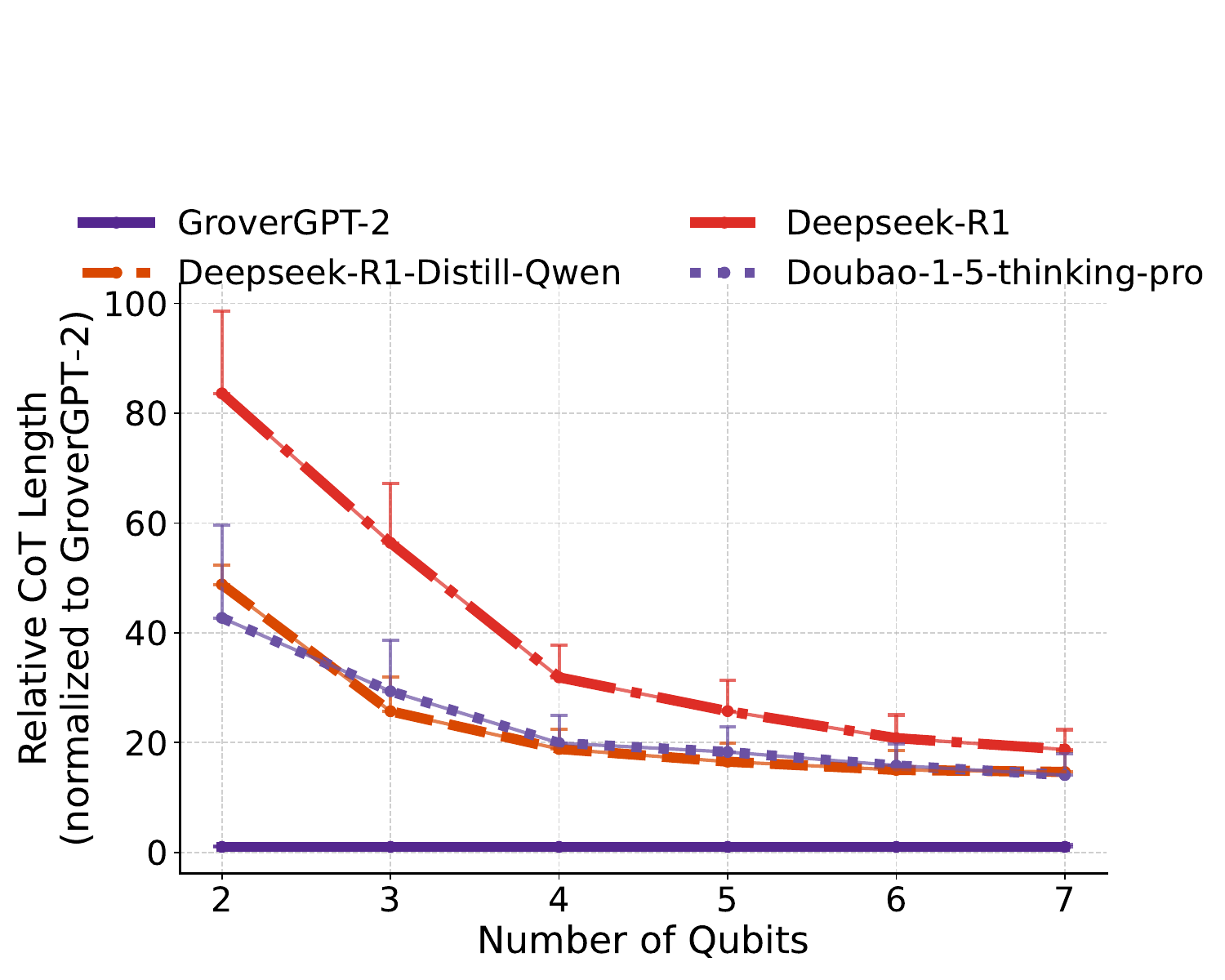}
\vspace{-0.5cm}
\caption{Comparison of the CoT lengths generated by different LLMs for the symbolic analysis of Grover's algorithm across qubit numbers ranging from 2 to 7. The values are normalized with GroverGPT+'s CoT length set as the baseline (1.0) for each qubit count. Solid lines show means; discrete error bars indicate uncertainty (mean $\pm$\, std).}
\label{fig:CoT_length}
\end{figure}

In this section, we analyze and compare the CoT lengths produced by various LLMs when performing symbolic analysis of Grover's algorithm. The objective is to investigate the computational overhead incurred by different models, particularly since longer CoT sequences can lead to increased latency and inference cost. For this analysis, we collected model-generated CoTs for each circuit and measured their lengths across qubit sizes. Each model’s average CoT length was then normalized by GroverGPT+’s CoT length at the same qubit count to provide a consistent relative comparison.

As illustrated in Supplementary Figure~\ref{fig:CoT_length}, GroverGPT+ consistently generates substantially shorter CoT reasoning sequences compared to baseline models. For example, at 2 qubits, GroverGPT+ maintains a normalized CoT length of 1.00, while baseline models often produce sequences that are dozens of times longer—some exceeding 40$\times$ or even 80$\times$ the length. This trend persists across higher qubit counts as well. Even at 7 qubits, GroverGPT+ maintains concise outputs, whereas other models continue to generate significantly longer CoTs, reflecting less focused or more verbose reasoning processes.


We can conclude that even at small qubit sizes, baseline models often require excessively long CoTs to reach a potentially valid answer. 
This indicates that, without targeted supervision, these models tend to produce verbose and sometimes redundant reasoning chains, reflecting weaker task alignment.
Moreover, as evidenced in Figure 3 in the main text, these long CoTs do not guarantee reliable performance, as baseline models still struggle to achieve consistent simulation accuracy, reinforcing that longer reasoning is not necessarily better in this context. 
By contrast, GroverGPT+’s compact and task-aligned CoTs result from explicit fine-tuning that leverages the structural logic of Grover’s algorithm, yielding both faster and more reliable outcomes.


\begin{figure} 
\centering
\includegraphics[width=0.7\textwidth]{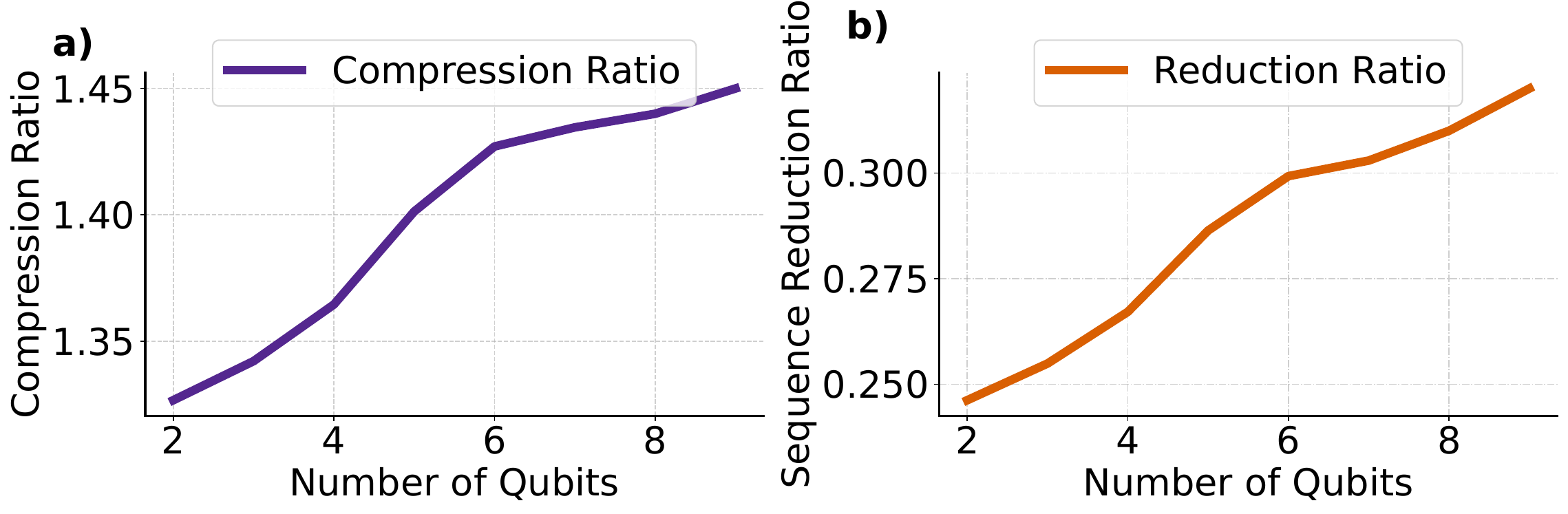}
\caption{
    Compression Ratio (a) and Sequence Reduction Ratio (b) across different number of qubits, showing the efficiency of quantum-native tokenization for token compression.
}
\label{fig:quantum_metrics}
\end{figure}

\section{Empirical Study of Quantum-Native Tokenization}
\label{appendix:Empirical Study of Quantum-Native Tokenization}

In this study, we investigate how the quantum-native tokenization performs under different number of qubits. This study is conducted following the below steps:

Firstly, we collect all the QASM circuit descriptions for Grover's algorithm under different numbers of qubits. For each given number of qubits $n$, there are $M_n$ QASM circuit descriptions. Secondly, we tokenize each QASM description using the base tokenizer of the LLaMA-3 model and our designed quantum-native tokenizer to obtain the length of each tokenized sequence. Specifically, for the $i$-th QASM circuit under $n$ qubits, let $L_{base}^{(n,i)}$ denote the sequence length obtained by the base tokenizer, and let $L_{quantum}^{(n,i)}$ denote the sequence length obtained by the quantum-native tokenizer. Thirdly, we calculate the compression ratio and the sequence reduction ratio for each number of qubits by averaging over all tokenized sequences under the same number of qubits. They can be formally defined as follows:

\begin{equation}
    \text{Compression Ratio}_n = \frac{1}{M_n}\sum_{i=1}^{M_n}\frac{L_{base}^{(n,i)}}{L_{quantum}^{(n,i)}}
\end{equation}

\begin{equation}
\begin{aligned}
    \text{Sequence Reduction Ratio}_n &= \frac{1}{M_n}\sum_{i=1}^{M_n}\frac{L_{base}^{(n,i)} - L_{quantum}^{(n,i)}}{L_{base}^{(n,i)}}
\end{aligned}
\end{equation}


As illustrated in Supplementary Figure~\ref{fig:quantum_metrics}, both the results measured using the compression ratio and the sequence reduction ratio indicate a higher computational efficiency as the number of qubits increases. Specifically, the compression ratio increases from below 1.35 to above 1.40, indicating that the quantum-native tokenizer consistently produces shorter token sequences compared to the LLaMA-3 base tokenizer, with the advantage becoming more prominent for larger circuits. Correspondingly, the sequence reduction ratio increases from below 0.26 to above 0.30, meaning that the token length saved by the quantum-native tokenizer becomes more substantial as the circuit complexity grows. This trend can be attributed to the structural design of the quantum-native tokenizer, which recognizes entire gate operations (e.g., \texttt{cx q[0],q[1];}) and qubit identifiers as cohesive, semantic units, whereas the LLaMA-3 tokenizer often splits them into meaningless subwords based on natural language rules. As Grover circuits scale with more qubits, the QASM descriptions exhibit increasingly repetitive or patterned structures (e.g., repeated oracle and diffuser subroutines). The quantum-native tokenizer is able to capitalize on these recurring syntactic forms and compress them more efficiently.

These results demonstrate the effectiveness of the proposed tokenizer in reducing sequence length, which directly leads to reduced memory and computational requirements during downstream model fine-tuning or inference. The shorter token sequences not only enable better GPU memory utilization but also reduce the context fragmentation problem inherent in sequences produced by the base tokenizer, particularly beneficial when handling circuits of increasing size and complexity.

\begin{figure*}[t]
\centering
\includegraphics[width=0.8\textwidth]{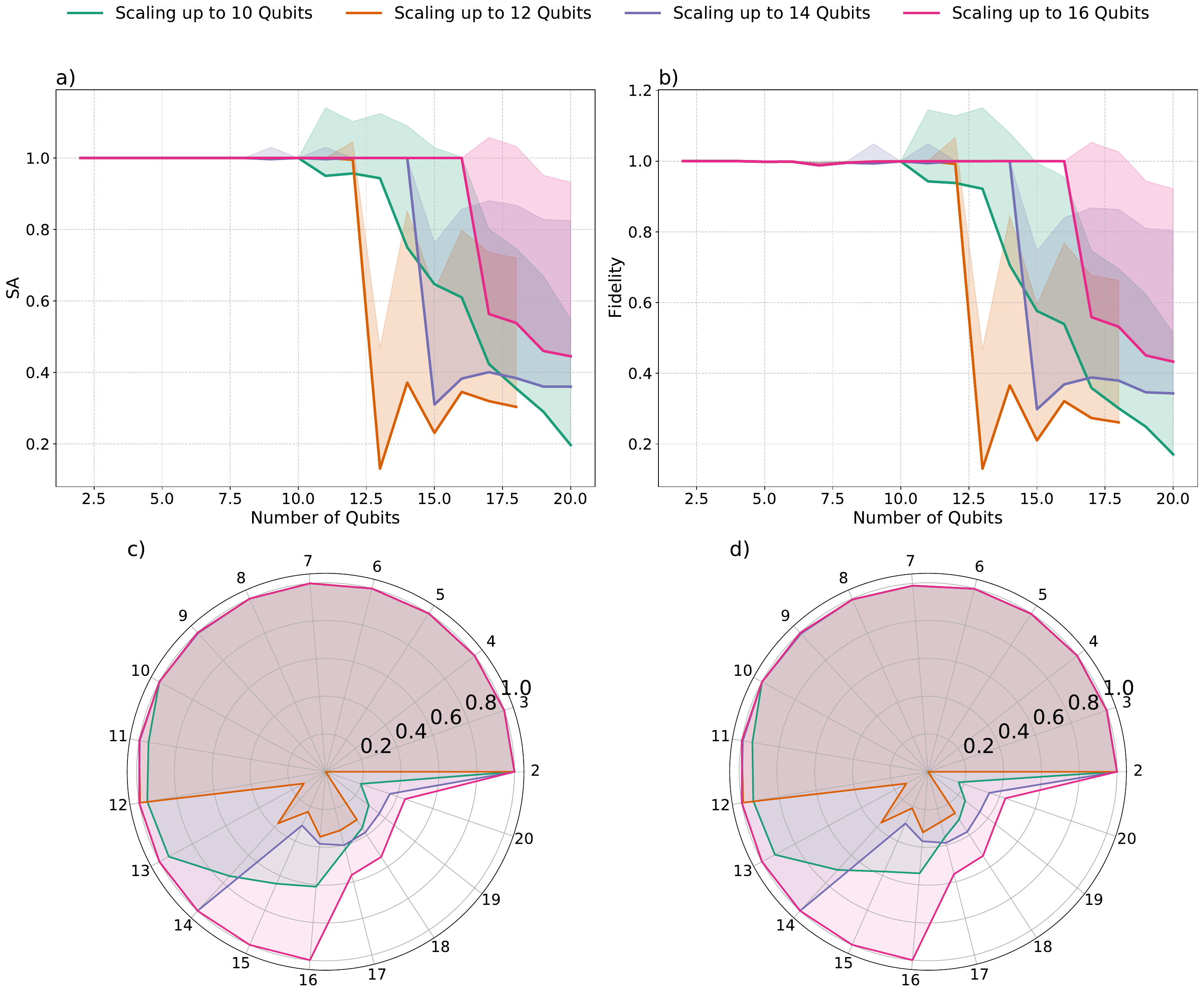}
\vspace{-0.3cm}
\caption{The extrapolation performance of GroverGPT+ under different training data scaling, including the scaling up to 10/12/14/16 qubits.}
\label{fig:Oracle_scaling_all}
\end{figure*}

\section{Extrapolation Performance of GroverGPT+}
\label{appendix:Scaling_law}


In this study, we investigate how GroverGPT+ performs when extrapolated to larger problem size, specifically as a function of the number of qubits.
Specifically, we investigate how the model's performance scales both within the training distribution and when generalizing to unseen problem sizes. Due to the excessive length of full-circuit inputs, which restricts the number of qubits to at most 10, a meaningful evaluation of the extrapolation cannot be reliably observed using data with full-circuit input. Therefore, we focus on the Oracle-only input setting. In particular, we construct four different training sets where the circuits with Oracle-only input cover qubit ranges spanning from 2 to 10, 12, 14 and 16 qubits, respectively. For each setting, we evaluate the model's performance across test circuits ranging from \(2\) to \(20\) qubits, measuring its ability to interpolate within the training distribution and extrapolate beyond it. All performance claims are empirically grounded in the tested ranges. 

The corresponding results are depicted in Supplementary Figure~\ref{fig:Oracle_scaling_all}, which reveal the model's performance behavior under different training ranges. For this figure, we use continuous shaded bands as error bars. Our key observations and corresponding analysis are as follows:

\begin{itemize}
    \item Firstly, the model demonstrates consistently high performance within the training distribution across all experimental conditions. Specifically, as evident in all figures, the evaluation metrics, including the SA and fidelity, reach or closely approach 1.0 within the qubit range included in the training datasets. This confirms that, within the bounds of the trained problem sizes, GroverGPT+ effectively learns to accurately analyze the quantum circuits and infer their outcomes.
    \item Secondly, when the training dataset is relatively small (e.g., covering qubits from 2 to 10), the model exhibits a noticeable degree of generalization capability to unseen data points. For instance, Supplementary Figure~\ref{fig:Oracle_scaling_all} (c, d) show that the model maintains consistently high performance in both SA (approximately 0.95) and fidelity (approximately 0.93) for qubit sizes of 11, 12, and 13, with only minor fluctuations across these larger circuit scales. However, a clear decreasing trend emerges as the circuit complexity grows beyond this range (14 to 20 qubits). Both SA and fidelity show parallel declining trends, suggesting that the generalization capability progressively deteriorates as the problem size increases. Upon inspecting specific output examples (as illustrated in Supplementary Note 15), we observed cases indicating errors in identifying the correct marked states, which likely contributes significantly to the observed decline in both metrics.
    \item Thirdly, when the dataset size increases further (covering ranges from 2 to 12 and 2 to 14), the high performance within the training range persists. Nevertheless, the generalization ability sharply decreases for larger qubit sizes outside these training distributions (13–20 and 15–20 qubits, respectively). The deterioration patterns for both SA and fidelity also closely resemble each other, suggesting a common underlying issue as above. However, when scaling the training range to 16 qubits, the generalization performance can surpass other ranges, indicating potential advantage of scaling the training size to larger qubit ranges. Similarly, we examine some model outputs, which reveal frequent inaccuracies in marked state reasoning, as exemplified in (as illustrated in Supplementary Note 15). 
\end{itemize}

While inspecting some failed cases similar to Supplementary Note 15, we find that GroverGPT+ performs well on instances involving single or two marked states. However, it fails to correctly identify the marked states when three marked states are required, primarily due to the omission of key reasoning steps in the CoT process. A plausible reason for the decreased generalization could be twofold: on one hand, the inherent increase in the complexity of quantum state representation and inference tasks as the number of qubits and marked states expands beyond the training distribution imposes greater demands on the model's learned representations; on the other hand, the phenomenon of catastrophic forgetting~\citep{mccloskey1989catastrophic, ratcliff1990connectionist} may also contribute, whereby exposure to an enlarged training range (i.e., circuits with higher qubit numbers) potentially disrupts the model’s previously acquired capabilities for smaller circuit sizes, thereby impairing its ability to generalize to unseen configurations.

\begin{tcolorbox}[
    colback=yellow!10,     
    colframe=black,        
    coltitle=white,        
    colbacktitle=black,    
    title=Supplementary Note 15: An example of output that misses key CoT steps, 
    fonttitle=\bfseries,   
    sharp corners,         
    boxrule=0.8pt,
    label=scaling_law_fail
]

=== Analysis === \\

The Oracle entity is extracted below: \\

... \textcolor{green}{(Correct)} \\

=== Block 1 ===\\
Operation sequence:\\
... \textcolor{green}{(Correct)} \\

State construction: \\
1. x \_gate\_q\_0: Present → 0, \textcolor{orange}{then → 0} \textcolor{red}{(Missing)} \\ 
2. x \_gate\_q\_1: Absent → 1, \textcolor{orange}{then → 10} \textcolor{red}{(Missing)} \\ 
3. x \_gate\_q\_2: Present → 0, \textcolor{orange}{then → 010} \textcolor{red}{(Missing)} \\ 

...

Final state: 00100000 \textcolor{red}{(Incorrect)}\\

...
\end{tcolorbox}


\section{Hyperparameter Settings}
\label{sec:hyperparameters}

In this section, we detail the key hyperparameters used for training GroverGPT+ in our experiments. The model was fine-tuned using LoRA with the configurations in Supplementary Table~\ref{tab:hyperparams_scope}:

\begin{table}[H]
\centering
\caption{Hyperparameter Scope for GroverGPT+}
\label{tab:hyperparams_scope}
\begin{tabular}{|c|c|c|}
\hline
\textbf{Hyperparameters} & \textbf{Value/Setting} & \textbf{Type} \\ 
\hline
Base Model & Llama-3-8B-Instruct & Fixed* \\
Optimizer & AdamW & Fixed \\
Learning Rate $\eta$ & \(2 \times 10^{-5}\) & Fixed \\
Batch Size $B$ & 1 & Fixed* \\
Gradient Accumulation Steps & 8 & Fixed \\
Effective Batch Size & 8 & Derived \\
Training Epochs $E$ & 10 & Fixed \\
Learning Rate Schedule & Cosine & Fixed \\
Warmup Steps & 20 & Fixed \\
Weight Decay & 0.0 & Fixed \\
Max Sequence Length & 4000 & Fixed \\
FP16 Mixed Precision & Enabled & Fixed \\
LoRA Target Modules & \{\textit{q\_proj}, \textit{v\_proj}\} & Fixed \\
LoRA Rank $r$ & 8 (default) & Fixed† \\
LoRA Alpha $\alpha$ & 32 (default) & Fixed† \\
Evaluation Strategy & Every 50 steps & Fixed \\
Evaluation Split & 10\% & Fixed \\
Max Gradient Norm & 1.0 & Fixed \\
Random Seed & 42 & Fixed \\
Dataset Shuffling & Disabled & Fixed \\
\hline
\end{tabular}
\end{table}

\noindent\footnotesize{
* Effective batch size calculated as $B_{\text{effective}} = B \times \text{gradient\_accumulation\_steps}$ \\
† Default values from LLaMA-Factory~\citep{zheng2024llamafactory} implementation
}

\clearpage
\bibliographystyle{apsrev4-2}
\bibliography{main}

\end{document}